\documentclass[10pt, journal]{./sty/IEEEtran}

\ifCLASSINFOpdf
  \usepackage[pdftex]{graphicx}
  \graphicspath{{./fig/}}
  \DeclareGraphicsExtensions{.pdf,.jpeg,.png}
\else
  \usepackage[dvips]{graphicx}
  \graphicspath{{./fig/}}
  \DeclareGraphicsExtensions{.eps}
  \fi

\usepackage[utf8]{inputenc} 
\usepackage[english]{babel} 
\usepackage{scrhack} 
\usepackage[scr=euler]{mathalfa} 
\usepackage{amssymb}
\usepackage{amsmath}
\usepackage{algorithm, algpseudocode}
\usepackage{trsym}
\usepackage{color}

\usepackage{emptypage}

\usepackage{pdflscape}
\usepackage{psfrag}
\usepackage{pstool}

\usepackage{subcaption}
\usepackage{mathtools}

\graphicspath{{./fig/}}
\DeclareGraphicsExtensions{.eps}
\usepackage{pgfplots}
\pgfplotsset{compat=newest}
\pgfplotsset{plot coordinates/math parser=false}
\usepackage{tikz}
\tikzset{>=latex}
\pgfsetxvec{\pgfpoint{1mm}{0}}
\pgfsetyvec{\pgfpoint{0}{1mm}}

\usetikzlibrary{patterns,plotmarks}
\usepackage{epstopdf}
\usepackage{accents}
 \allowdisplaybreaks

\usepackage{bm, upgreek}
\newcommand*{\lambdafHl}{{\mf{\Lambda}_h^{(l)}}}
\newcommand*{\pil}{p}
\newcommand*{\upmuvec}{{\bm{\upmu}}}

\newcommand{\ve}{\mathbf}
\newcommand{\m}{\mathbf}
\newcommand{\veel}[1]{#1} 
\newcommand*{\mel}[1]{\left[#1\right]}

\newcommand{\mf}[1]{\mathbf{\tilde{\mathbf{#1}}}} 
\newcommand{\vef}[1]{\mathbf{\tilde{\mathbf{#1}}}} 
\newcommand{\vefel}[1]{\tilde{#1}} 

\newcommand{\est}[1]{\hat{#1}}

\newcommand*{\elem}[1]{\left[ #1 \right]}

\newcommand*{\E}[1]{\mathrm{E}\left\{ #1 \right\}}
\newcommand*{\diag}[1]{\mathrm{diag}\left( #1 \right)}
\newcommand*{\tr}[1]{\mathrm{tr}\left( #1 \right)}

\newcommand*{\e}{\mathrm{e}}

\newcommand*{\cfo}{\text{CFO}}

\newcommand*{\ici}{\text{ICI}}

\newcommand*{\cp}{\text{cp}}
\newcommand*{\uw}{\text{uw}}

\newcommand*{\delayspread}{\tau_\text{RMS}}

\newcommand*{\downs}{{pl}}

\newcommand*{\noise}{\ve{n}}
\newcommand*{\varnoise}{\sigma_n^2}

\newcommand*{\noisef}{\ve{v}}
\newcommand*{\varnoisef}{\sigma_v^2}

\newcommand*{\nElemNoise}{\veel{n}[u]}  

\newcommand*{\noisefpil}{\ve{v}''}  
\newcommand*{\kElemNoisePil}{\veel{v}''[k]}  
\newcommand*{\kElemNNoisePil}{\veel{v}'''[k]}  

\newcommand*{\noiseffull}{\ve{v}'}  
\newcommand*{\kElemVNoisefull}{\veel{v}'[k]}  

\newcommand*{\varphil}{\varphi_l}
\newcommand*{\varphilest}{\est{\varphi}_l}
\newcommand*{\psil}{\psi_l}

 \newcommand*{\lambdaNl}{\m{\Lambda}'^{(l)}}
\newcommand*{\lambdaNfl}{\mf{\Lambda}'^{(l)}}
\newcommand*{\lambdafl}{{\mf{\Lambda}^{(l)}}}

\newcommand*{\Yrxl}{\ve{y}_r^{(l)}}  

\newcommand*{\nElemYrxl}{\veel{y}_r^{(l)}[u]}  

\newcommand*{\Yfrxl}{\vef{y}_r^{(l)}}
\newcommand*{\kElemYfrxl}{\vefel{y}_r^{(l)}[k]}  

\newcommand*{\Yfdown}{\vef{y}_{\downs}}
\newcommand*{\Yfdownl}{\vef{y}_{\downs}^{(l)}}

\newcommand*{\Xfl}{\vef{x}''^{(l)}}  

\newcommand*{\mElemXfhl}{\vefel{x}_h^{(l)}[m]}  
\newcommand*{\kElemXfhl}{\vefel{x}_h^{(l)}[k]}  

\newcommand*{\mElemXfl}{\vefel{x}''^{(l)}[m]}  
\newcommand*{\kElemXfl}{\vefel{x}''^{(l)}[k]}  

\newcommand*{\Xfdownl}{\vef{x}_d^{(l)}}

\newcommand*{\datal}{\ve{d}^{(l)}}

\newcommand*{\IpkElemUW}{\vefel{x}_u[i_{p,k}]}

\newcommand*{\dataicil}{\ve{d}_\ici^{(l)}}
\newcommand*{\kElemdataicil}{\veel{d}_\ici^{(l)}[k]}

\newcommand*{\idxn}{u}

\newcommand*{\pestl}{\est{\ve{p}}^{(l)}}
\newcommand*{\kElempestl}{\est{\veel{p}}^{(l)}[k]}
\newcommand*{\kElempicil}{\veel{p}^{(l)}_\ici[k]}

\newcommand*{\kElemp}{\veel{p}[k]}
\newcommand*{\mElemp}{\veel{p}[m]}

\newcommand*{\kElemwp}{\veel{w}_p[k]}

\newcommand*{\kElemICI}{\veel{i}^{(l)}[k]}  

\DeclareMathOperator*{\argmin}{argmin}

\renewcommand{\baselinestretch}{0.98}

\begin{document}

\title{On the Inclusion and Utilization of Pilot Tones in Unique Word OFDM}

\author{Christian
  Hofbauer,~\IEEEmembership{Member,~IEEE}, Werner
  Haselmayr,~\IEEEmembership{Member,~IEEE}, Hans-Peter
  Bernhard,~\IEEEmembership{Senior Member,~IEEE}, and Mario~Huemer,~\IEEEmembership{Senior Member,~IEEE}
  \thanks{The authors are with the Silicon Austria Labs GmbH, Linz, Austria
  (e-mail: \{christian.hofbauer, hans-peter.bernhard\}@silicon-austria.com), with the
  Institute of Signal Processing,
Johannes Kepler University Linz, Linz, Austria (e-mail:
mario.huemer@jku.at), and with the Institute for Communications Engineering and
RF-Systems, Johannes Kepler University Linz, Linz, Austria (e-mail:
werner.haselmayr@jku.at), respectively.

This work has been supported by Silicon Austria Labs (SAL), owned by the
Republic of Austria, the Styrian Business Promotion Agency (SFG), the federal
state of Carinthia, the Upper Austrian Research (UAR), and the Austrian
Association for the Electric and Electronics Industry (FEEI).
  }
}


\maketitle

\begin{abstract}
Unique word-orthogonal frequency division multiplexing (UW-OFDM) is known to
provide various performance benefits over conventional OFDM using cyclic
prefixes (CP). Most important, UW-OFDM features excellent spectral sidelobe
suppression properties and an outstanding bit error ratio performance.
Current research has mainly focused on principle performance
bounds of UW-OFDM, with less attention on challenges aside from idealized
communication scenarios, such as system
parameter estimation tasks. In this work we present an
approach for including frequency pilots tones into the UW-OFDM signaling scheme, which
can then be utilized for these estimation tasks. Suitable optimization criteria
are presented and interactions of pilots with data symbols are highlighted. Pilot
tone based estimation of a carrier frequency offset (CFO) is conducted as an estimation example, revealing considerable differences to conventional
OFDM. Simulation results in a multipath environment demonstrate
a significantly increased estimation accuracy in UW-OFDM over CP-OFDM, which
becomes even more dominant with an increasing CFO. This performance difference is due to the inherent redundancy present in an UW-OFDM signal.
\end{abstract}


\IEEEpeerreviewmaketitle

\section{Introduction}

The Unique Word (UW)-OFDM signaling scheme introduced in \cite{Huemer10_1} uses a
deterministic sequence in the guard interval instead of the conventional cyclic
prefix (CP). The introduction of the unique word within the discrete Fourier transform~(DFT)
interval requires the introduction of redundancy in the frequency domain. This
redundancy can advantageously be utilized to provide several beneficial properties, such
as superior spectral shaping characteristics
\cite{Huemer12_1,Rajabzadeh13,Rajabzadeh14,Rajabzadeh18,Lang19} or outstanding bit error ratio (BER)
performance for linear \cite{Huemer2011,Huemer11_1, Huemer12_1,Hofbauer12,Huemer13,Hofbauer16_1}, non-linear
\cite{Huemer12_2,Onic14,Onic13} as well as iterative \cite{Haselmayr14,Haselmayr19}
receivers\footnote{In this context, the term \emph{iterative receiver} refers to an iterative
  exchange of realibility (soft) information between detector and decoder \cite{Douillard95, Tuechler02_1}.}. 
 
Various other approaches labeled KSP-OFDM (known symbol padding) \cite{Welden08}, TDS-OFDM (time domain synchronous) \cite{China06,Ong10,Tang07},
PRP-OFDM (pseudorandom prefix) \cite{Muck06}, OFDM with PN (pseudo noise) sequence \cite{Tang07} or
even OFDM with UW \cite{Jingyi02} implement deterministic sequences in the
guard interval. Differing from each other in the specific instance of the
sequence, all those schemes implement the guard interval outside of the DFT interval. Hence,
no redundancy as in UW-OFDM is present, precluding thus the advantageous properties of UW-OFDM signals.
 
So far, UW-OFDM has been investigated regarding its principle performance bounds, implying the assumption of
several idealized conditions, such as perfect timing, carrier phase or carrier frequency synchronization. The investigation of real-world aspects has been
limited to computational complexity analyses \cite{Huemer11_1,Onic11}, peak to
average power ratio (PAPR) and peak to minimum power ratio (PMR) considerations
\cite{Huber12_1,Rettelbach12}, as well as the effects of channel estimation
errors on the BER performance \cite{Huemer12_1}. 

One important task in all
real-world communication systems is the estimation of various system
parameters, which is carried out in time, frequency or both
domains. As such, a deterministic time domain sequence like a UW can
already offer valuable contributions. However, the domain actually applicable and
suitable for a particular estimation problem usually depends on various aspects. These aspects may on the one hand encompass specific requirements of each problem
regarding accuracy, computational complexity or real-time constraints, or on
the other hand prerequisites like e.g., the underlying signaling scheme.
For instance,
multicarrier schemes as e.g., OFDM may  --- due to their inherent structure
--- prefer utilizing pilot symbols in the frequeny domain (also known as pilot
tones) to conduct such system parameter estimation tasks \cite{Classen94}. 

This work is based on \cite{Hofbauer16} and extends the UW-OFDM concept in order to enable the inclusion of
deterministic pilot symbols at dedicated subcarriers in the frequency
domain. We therefore provide a generic
UW-OFDM signaling framework that enables pilot symbols in both time (i.e., UWs)
and frequency domain (i.e., pilot tones). However, the utilization of UWs for actual
estimation tasks is beyond the scope of this work\footnote{In fact, a
  utilization of UWs for estimation tasks would limit the comparability with
  CP-OFDM.}, and the interested reader is e.g.,
referred to \cite{Welden08,Tang07,Muck06,Ehsa201909,Ehsa202004} for details. Instead, we show that due to its specific structure, the UW-OFDM
signaling scheme provides a better estimation performance than conventional
CP-OFDM, even for an estimator solely based on pilot tones.
In this context, we address the following aspects. We investigate the inclusion of pilot tones and study their interaction with data symbol transmission. We
introduce suitable cost functions for the optimization of the UW-OFDM transmit
signal, identify potential optimization parameters and study their impact on
the cost functions. Exemplary generator matrices are presented and their
similarities with matrices from pilotless UW-OFDM systems are evaluated. Further, an
UW-OFDM signaling model incorporating a carrier frequency offset~(CFO) is
provided. Moreover, we present a pilot tone based common phase error~(CPE) estimation technique, elaborate on the differences to conventional OFDM systems and compare their estimation performance.


We note that this paper expands its conference version \cite{Hofbauer20_1} in
several directions. We describe the generic signaling framework for pilot
symbol insertion in more
detail. Further, we thoroughly develop the CFO model step-by-step until finally
yielding the signaling model presented in \cite{Hofbauer20_1}. We investigate the approximation error in the signaling model due to partially neglecting intercarrier interference (ICI) and justify its neglection. Moreover, the pilot tone based estimator for the CPE utilized in
both works is also explained in more detail, and complemented
by an approach to derive the CFO based on the CPE. Finally, the performance of
the pilot tone based estimator is evaluated for two different generator matrices instead of only one.

\textit{Notation:} Vectors and
matrices are denoted in bold face lower case $\ve{a}$ and upper
case letters $\m{A}$, respectively. A tilde is used to explicitly label
variables in the frequency domain ($\vef{a}$, $\mf{A}$). The $k$th element of a vector $\ve{a}$ is named $a[k]$, $[\m{A}]_{k,l}$ addresses the
element in column $k$ and row $l$, $[\m{A}]_{k,*}$ represents all
elements of row number $k$, and $[\m{A}]_{*,l}$  all
elements of column number $l$. The transpose operation is expressed as
$(\cdot)^T$, the conjugate transpose or
Hermitian as $(\cdot)^H$, expectation as $\E{\cdot}$, $\tr{\m{A}}$ denotes
the trace operation, $\diag{\m{A}}$ extracts the main diagonal entries
of a matrix $\m{A}$, and $(\cdot)^\dagger$ corresponds to the Moore-Penrose
Pseudo-Inverse. The identity matrix is given by $\m{I}$ and a zero matrix as
$\m{0}$. A vector $\ve{a}\sim\mathcal{CN}\left(\upmuvec,\m{C}\right)$
denotes a circularly symmetric complex Gaussian noise vector with mean
$\upmuvec$ and covariance matrix $\m{C}$. Further, $\hat{\ve{a}}$ corresponds to an estimate of $\ve{a}$.
An underlined letter shall emphasize the \underline{m}otivation behind the
specific subscript (or superscript) of $\ve{a}_m$. For all signals and systems in this work, the usual equivalent complex baseband
representation applies. 

The remainder of this paper is organized as follows. Sec.~\ref{sec:review}
briefly recaps the UW-OFDM signaling scheme, which is then extended by
frequency domain pilot tone insertion in Sec.~\ref{sec:pilots}. In Sec.~\ref{sec:cfo_phase_rotation} we investigate the
utilization of pilot tones for estimating the CPE and CFO. Subsequently, Sec.~\ref{sec:performance} compares the estimation
performance in UW-OFDM against CP-OFDM, and Sec.~\ref{sec:conclusion}
finally concludes our work.

\section{UW-OFDM Signaling Scheme}\label{sec:review}
The UW-OFDM concept is briefly reviewed in the following, a detailed analytical
derivation can be found in \cite{Huemer14}.
\subsection{Transmit Symbol Generation}\label{sec:txsymb_generation}
\begin{figure}[hbt]
\centering
\begin{tikzpicture}
	\clip (-25, -7) rectangle (50, 17);
	\draw (0, 0) rectangle +(30, 8);
	\draw[pattern=horizontal lines light gray] (30, 0) rectangle +(10, 8);
	\node at (35, 4){UW};
	\draw[|<->|] (30, 12) -- node[above] {$N_g$} +(10, 0);
	\draw[|<->|] (0, 10) -- node[above] {$N$} +(40, 0);
	\draw (-40, 0) rectangle +(30, 8);
	\draw[pattern=horizontal lines light gray] (-10, 0) rectangle +(10, 8);
	\node at (-5, 4){UW};
	\draw (40, 0) rectangle +(30, 8);
	\draw[|<->|] (0, -2) -- node[below] {UW-OFDM symbol} +(40, 0);
\end{tikzpicture}
\caption{Structure of an UW-OFDM symbol.}
\label{fig:uwofdm_symbol}
\end{figure}
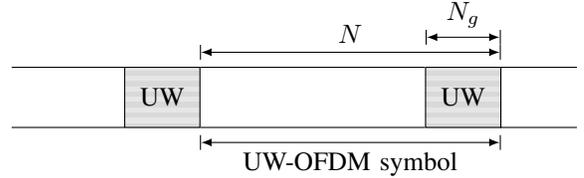

Let $\ve{x}_u\in\mathbb{C}^{N_u \times 1}$ be a predefined sequence which we call
\underline{U}W. This unique word shall form the tail of each OFDM time domain
symbol vector of length $N$ and occupy the \underline{g}uard interval of equal length
$N_g=N_u$, as illustrated in Fig.~\ref{fig:uwofdm_symbol}.
Hence, an OFDM time domain symbol $\ve{x}'\in \mathbb{C}^{N
  \times 1}$ consists of two parts and is of the form
  $\ve{x}'=\begin{bmatrix}\ve{x}_{pl}^T & \ve{x}_u^T \end{bmatrix}^T$, 
whereas $\ve{x}_{pl}\in\mathbb{C}^{(N-N_u) \times 1}$ carries the
\underline{p}ay\underline{l}oad affected by the data symbols. Energy arguments elaborated
in \cite{Huemer14,Onic10} suggest to first generate an OFDM time domain symbol with a
zero-word
$\ve{x} = \begin{bmatrix}\ve{x}_{pl}^T & \ve{0}^T\end{bmatrix}^T$, 
followed by adding the desired UW in time domain in a second step to obtain
$\ve{x}' = \ve{x} + \begin{bmatrix}\ve{0}^T & \ve{x}_u^T\end{bmatrix}^T$. 
As in conventional OFDM, unused zero subcarriers together with $N_d$ data symbols
$\ve{d}\in\mathcal{A}^{N_d \times 1}$ drawn from a symbol alphabet $\mathcal{A}$
 shall form an OFDM symbol $\vef{x}\in\mathbb{C}^{N \times
  1}$ in the frequency domain. Since UW-OFDM additionally demands a
zero-word in time domain as part of $\ve{x}$, the system of
equations
$\m{F}_N^{-1}\vef{x}=\ve{x}$,
whereas $\m{F}_{N}^{-1}$ denotes the inverse DFT of size $N$ with $\m{F}_{N}^{-1} = \frac{1}{N}\m{F}_{N}^H$ and $\mel{\m{F}_N}_{k,l} = \e^{-j\frac{2\pi}{N}kl}$ , can only be fulfilled by reducing the
number of data symbols in the frequency domain by at least $N_u$, and instead
introducing a certain kind of \underline{r}edundancy. For this purpose, let
$\m{G}\in\mathbb{C}^{(N_d+N_r)\times N_d}$ be a generator matrix with
$N_r=N_u$, which ensures all these requirements. Furthermore, and
w.l.o.g., we decompose $\m{G}$ into submatrices according to
\begin{equation}
\m{G}=\m{A}\begin{bmatrix}\m{I} \\ \m{T} \end{bmatrix}\label{equ:tx016}
\end{equation}
to distinctly address the different requirements on $\m{G}$. Hence, this leads to the time-frequency relation
\begin{equation}
  \ve{x}=\m{F}_N^{-1}\m{B}\m{A}\begin{bmatrix}\m{I} \\ \m{T} \end{bmatrix}\ve{d}=\begin{bmatrix}\ve{x}_{pl}\\ \ve{0}\end{bmatrix}.\label{equ:tx011}
\end{equation}
Independent of $\m{G}$, matrix $\m{B}\in\{0,1\}^{N \times (N_d+N_r)}$ models the
insertion of optional $N_z$ zero subcarriers for spectral shaping reasons,
completing the total number of subcarriers of $N=N_d+N_r+N_z$.

Within $\m{G}$, the identity matrix $\m{I}$ takes care of
mapping the data symbols onto $\ve{x}$. Further, $\m{T}$ automatically
generates the desired zero-word in time domain by simply choosing $\m{T} =
-(\m{M}_{22})^{-1}\m{M}_{21}$, $\m{T}\in\mathbb{C}^{N_r\times N_d}$, with
appropriately sized submatrices  from $\m{F}_N^{-1} \m{B} \m{A}= \left[\begin{smallmatrix} \m{M}_{11} & \m{M}_{12}
    \\ 			\m{M}_{21} & \m{M}_{22}\end{smallmatrix}\right]$, regardless of
the specific realizations of $\m{B}$ and $\m{A}$. Finally, $\m{A}\in\mathbb{C}^{(N_d+N_r)\times (N_d+N_r)}$ provides the degrees of freedom
to optimize the UW-OFDM generator matrix towards an appropriate cost
function to yield certain desired properties. In other words, the decomposition
of $\m{G}$ according to \eqref{equ:tx016} has transformed an originally
constrained optimization problem into an
unconstrained one, as $\m{T}$ will ensure fulfilling the zero-word constraint for any realization of $\m{A}$.
The only restriction on $\m{A}$ is its non-singularity due to the inversion of
$\m{M}_{22}$. Based on the specific design of $\m{A}$,
different classes of UW-OFDM systems emerge, such as \emph{systematically
  encoded} or \emph{non-systematically encoded} UW-OFDM \cite{Huemer14},
resulting in a huge collection of different systems with each individual providing benefits for a
different scenario\footnote{Due to space limitations, the different kinds of generator
  matrices are not detailed further, instead, the interested reader is
  referred to \cite{Hofbauer16} for a detailed discussion.}. 

\subsection{Receiver}
After the transmission over a dispersive channel, a received UW-OFDM symbol in the
frequency domain carrying the \underline{p}ay\underline{l}oad (zero subcarriers
are excluded) can be formulated as 
\begin{equation}
		\Yfdown = \mf{H}\m{G}\ve{d} + \mf{H}\m{B}^T \vef{x}_u + \m{B}^T \m{F}_N \ve{n}, \label{equ:rx001}
\end{equation}
whereas $\mf{H}\in\mathbb{C}^{(N_d+N_r)\times (N_d+N_r)}$ denotes
a diagonal channel matrix with the sampled channel frequency
response on its main diagonal, $\vef{x}_u=\m{F}_N \begin{bmatrix}
  \ve{0}^T &\ve{x}_u^T \end{bmatrix}^T$ corresponds to the frequency domain version
of the UW, and  $\ve{n}\in\mathbb{C}^{N \times 1}$ to a noise vector with
$\ve{n}\sim\mathcal{CN}\left(\ve{0},\sigma_n^2\m{I}\right)$. Subtracting the UW
  induced offset according to $\vef{y}=\Yfdown - \mf{H} \m{B}^T
  \vef{x}_u$ (assuming that the channel matrix $\mf{H}$ or at least an
  estimate of it is available) yields the linear model
\begin{equation}
		\vef{y} = \mf{H}\m{G}\ve{d} + \ve{v}, \label{equ:009}
\end{equation}
with a noise vector $\ve{v} = \m{B}^T \m{F}_N \ve{n}$,
$\ve{v}\sim\mathcal{CN}\left(\ve{0},N\sigma_n^2\m{I}\right)$. One possibility
to obtain a data estimate is to
apply a linear minimum mean square error (LMMSE) estimator
\begin{align}
\est{\ve{d}} =  (\m{G}^H\mf{H}^H \mf{H}\m{G} + \tfrac{N \sigma_n^2}{\sigma_d^2}\m{I})^{-1} \m{G}^H\mf{H}^H\vef{y},\label{equ:rx002}
\end{align}
given a zero-mean data vector with covariance matrix $\sigma_d^2 \m{I}$. The covariance matrix of the error $\ve{e}
= \ve{d}-\est{\ve{d}}$ is then
\begin{equation}
	\m{C}_{\veel{e}\veel{e}} = \E{\ve{e}\ve{e}^H}=N \sigma_n^2 (\m{G}^H\mf{H}^H \mf{H}\m{G} + \tfrac{N \sigma_n^2}{\sigma_d^2}\m{I})^{-1}.  \label{equ:rx003}
\end{equation}

\section{Inclusion of Pilot Tones in UW-OFDM}
\label{sec:pilots}
In this section the UW-OFDM signaling scheme is extended to allow the
inclusion of deterministic pilot symbols $\ve{p}\in \mathbb{C}^{N_p\times 1}$ at
dedicated subcarriers in the
frequency domain. The presented framework is based on partitioning the
signal $\ve{x}$ into two additive terms. The first
term maps the \underline{d}ata symbols on $\ve{x}$ by using a generator matrix
$\m{G}_d\in\mathbb{C}^{(N_d+N_r+N_p)\times N_d}$, and the second term
incorporates the \underline{p}ilot symbols by using
$\m{G}_p\in\mathbb{C}^{(N_d+N_r+N_p)\times N_p}$, yielding
\begin{align}
  \ve{x}=\m{F}_N^{-1}\m{B}\m{G}_d\ve{d} + \m{F}_N^{-1}\m{B}\m{G}_p\ve{p}= \begin{bmatrix} \ve{x}_d \\ \ve{0} \end{bmatrix}.\label{equ:pil003} 
\end{align}
Matrix $\m{B}\in\mathbb{C}^{N\times(N_d+N_r+N_p)}$ models the insertion of optional zero
subcarriers and exactly coincides with matrix
$\m{B}$ from the pilotless case in Sec.~\ref{sec:txsymb_generation}\footnote{Note that $\m{B}$ is identical for 
 the pilot based and the pilotless case, but the definition of the dimensions
 differ due to the additional parameter $N_p$. Contrary to
 Sec.~\ref{sec:txsymb_generation}, it now holds that $N-N_z=N_d+N_r+N_p$
 instead of $N-N_z=N_d+N_r$.}. 

The partitioning of the transmit signal in \eqref{equ:pil003} into two additive
and independent terms suggests that these two terms can be optimized
independently from each other as part of two distinct optimization problems
with distinct cost functions. These problems will be tackled in the following.


\subsection{Optimization of Data Dependent Term}\label{sec:opt_data_term}

\begin{figure}[!htb]
  \centering
  \begin{subfigure}[b]{0.35\textwidth}
    \hspace{0.3cm}\scalebox{0.5}{
%
\begin{tikzpicture}

\begin{axis}[%
width=4.047in,
height=3.566in,
at={(0.679in,0.481in)},
scale only axis,
point meta min=0,
point meta max=0.841553421417078,
axis on top,
xmin=0.5,
xmax=32.5,
y dir=reverse,
ymin=0.5,
ymax=52.5,
axis background/.style={fill=white},
legend style={legend cell align=left, align=left, draw=white!15!black},
colormap={mymap}{[1pt] rgb(0pt)=(1,1,1); rgb(64pt)=(0,0,0)},
colorbar
]
\addplot [forget plot] graphics [xmin=0.5, xmax=32.5, ymin=0.5, ymax=52.5] {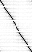};
\end{axis}
\end{tikzpicture}%
}
    \caption{$|\m{G}'_d|$}
    \label{fig:Gd1_abs}
    \vspace{0.5cm}
  \end{subfigure}

  \begin{subfigure}[b]{0.35\textwidth}
    \hspace{0.3cm}\scalebox{0.5}{
%
\begin{tikzpicture}

\begin{axis}[%
width=4.047in,
height=3.566in,
at={(0.679in,0.481in)},
scale only axis,
point meta min=0,
point meta max=0.404388521268797,
axis on top,
xmin=0.5,
xmax=32.5,
y dir=reverse,
ymin=0.5,
ymax=52.5,
axis background/.style={fill=white},
legend style={legend cell align=left, align=left, draw=white!15!black},
colormap={mymap}{[1pt] rgb(0pt)=(1,1,1); rgb(64pt)=(0,0,0)},
colorbar
]
\addplot [forget plot] graphics [xmin=0.5, xmax=32.5, ymin=0.5, ymax=52.5] {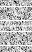};
\end{axis}
\end{tikzpicture}%
}
    \caption{$|\m{G}''_d|$}
    \label{fig:Gd2_abs}
  \end{subfigure}
  \caption{Magnitude of entries of data generator matrices $\m{G}_d$  based on
    the setup in Tab.~\ref{tab:setups}.}
  \label{fig:Gd_abs}
\end{figure}

The optimization of the data dependent term can be formulated as an optimization
problem for $\m{G}_d$ given as
\begin{align}
  \breve{\m{G}}_{d}=\argmin_{\m{G}_d}\left\{J_d\right\} \hspace{0.3cm} \mathrm{s.t.} \hspace{0.15cm} 		\m{F}_N^{-1}\m{B}\m{G}_d = \begin{bmatrix} \m{\Xi} \\ \m{0} \end{bmatrix}, \label{equ:pil004}
\end{align}
with $J_d$ denoting an appropriate cost function and
$\m{\Xi}~\in~\mathbb{R}^{N \times (N-N_u)}$ an arbitrary matrix. Similar to the
previous section, let us choose the approach
\begin{equation}
\m{G}_d=\m{B}_p\m{A}_d\begin{bmatrix}\m{I} \\ \m{T}_d \end{bmatrix},\label{equ:pil005}
\end{equation}
whereas
$\m{A}_d\in\mathbb{R}^{(N_d+N_r)\times(N_d+N_r)}$, $\m{B}_p\in\mathbb{C}^{(N_d+N_r+N_p)\times(N_d+N_r)}$,
and $\m{T}_d\in\mathbb{C}^{N_r \times N_d}$ with
$\m{T}_d=\m{M}'^{-1}_{22}\m{M}'_{21}$. Matrix $\m{B}_p$ places zeros at
the positions of the $N_p$ \underline{p}ilot
subcarriers and therefore ensures that the data part $\m{G}_d\ve{d}$ in
\eqref{equ:pil003} will not
be spread onto the pilot subcarriers. In contrast, however, the pilot part with
$\m{G}_p\ve{p}$ will overlay the data part to some extent to fulfill the zero
word constraint, which will be discussed later on in
Sec.~\ref{sec:opt_pilot_genmtx}. The submatrices
$\m{M}'_{21}\in\mathbb{C}^{N_u \times N_d}$ and 
$\m{M}'_{22}\in\mathbb{C}^{N_u \times N_r}$ with $N_r=N_u$ follow from
$\m{F}_N^{-1}\m{B}\m{B}_p\m{A}_d=\left[\begin{smallmatrix}\m{M}'_{11} &
    \m{M}'_{12}\\ \m{M}'_{21} & \m{M}'_{22} \end{smallmatrix}\right].$ For pilotless UW-OFDM systems, minimizing the sum of the error covariances
after data estimation at a fixed SNR turned out to be a well-chosen cost
function $J_d$ for
finding generator matrices. In \cite{Huemer12_1} we
showed that setting up $J_d$ assuming AWGN conditions with $\m{H}=\m{I}$, an approach
to yield an independence of $\m{G}_d$  from dedicated
channel realizations, delivers instances of $\m{G}_d$ that feature an outstanding BER performance compared to conventional OFDM. This is especially true for frequency-selective channels, which is
particularly astonishing due to the opimization towards an AWGN and thus frequency-flat channel.
Therefore, this motivates to apply the same
optimization criterion in case of UW-OFDM symbols with pilot
subcarriers. Considering \eqref{equ:pil003}, the transmit signal
$\ve{x}''$ can be modelled as
\begin{equation}
\ve{x}'' = \m{F}_N^{-1}\left(\m{B}\m{G}_d\ve{d} + \m{B}\m{G}_p\ve{p} +
\vef{x}_u\right). \label{equ:pil018}
\end{equation}
Similar
to \eqref{equ:rx001}, the frequency domain signal at the receiver (the zero
subcarriers are already excluded) can be expressed as
\begin{equation}
\Yfdown = \mf{H}\m{G}_d\ve{d} + \mf{H}\m{G}_p\ve{p} + \mf{H}\m{B}^T \vef{x}_u
+ \m{B}^T \m{F}_N \noise. \label{equ:pil019}
\end{equation}
Then, the known signal parts (assuming that
$\mf{H}$ or at least an estimate of it is available) caused by the UW and the
pilots are subtracted from $\Yfdown$, yielding the linear model
\begin{align}
\vef{y}&= \Yfdown - \mf{H}\m{B}^T\vef{x}_u - \mf{H}\m{G}_p\ve{p}\\
&= \mf{H}\m{G}_d\ve{d} +\noisef.\label{equ:pil020}
\end{align}
Putting differences in dimensionality and actual realizations of the
involved terms aside, the principle structure of a linear model is the same as in \eqref{equ:009}.
We thus conclude that every UW-OFDM approach --- regardless of the presence
of pilot symbols in the frequency domain --- leads to the same basic transmission model.
The main difference between an UW-OFDM system with and without pilot tones is
an additional subtraction of the pilot induced offset on top of the UW
part to yield the linear model. Based on \eqref{equ:pil020}, the following conclusions can
be drawn:
\begin{itemize}
\item The same structure of a linear transmission model for data symbols enables the deployment of the same
  estimator concepts as in the case without pilots.
\item The same structure of a linear model permits the same optimization procedure as in the
  pilotless case, which means the same cost function $J_d$, as well as
  the same tools to solve the optimization problem in
  \eqref{equ:pil004}. As a consequence, the steepest descent algorithm from
  \cite{Huemer12_1} developed for the pilotless case can be utilized without adaptations. 
\item Pilots are introduced for system parameter estimation purposes and then simply
  subtracted before the data estimation process, as they do not contribute any
  information to the latter. In this sense, the pilot dependent part can be designed
  independently from the data dependent part, which is discussed in detail in
  Sec.~\ref{sec:opt_pilot_genmtx}.  
%
\end{itemize}
Fig.~\ref{fig:Gd_abs} illustrates two exemplary generator matrices obtained from solving the optimization problem in
\eqref{equ:pil004} for the setup given in Tab.~\ref{tab:setups}. These
matrices  --- referred to as $\m{G}'_d$ and $\m{G}''_d$ ---   are the result of applying the steepest descent algorithm, initialized once with a
permutation matrix $\m{P}$ (derived from $\mathcal{I}_r$ in Tab.~\ref{tab:setups}) yielding $\m{A}^{(0)}=\m{P}$, and once with
$\mel{\m{A}^{(0)}}_{ij} \sim \mathcal{N}(0,1)$, respectively. The same
initialization approaches have been used for the pilotless case. Clearly, the similarities with $\m{G}'$
and $\m{G}''$ from the pilotless case in \cite{Huemer12_1} are immediately apparent,
only the setup differs slightly due to the additional $N_p$ pilot
subcarriers. $\m{G}_d'$ maps, analogeously to $\m{G}'$, a symbol mainly onto one subcarrier,
thus behaving much like a conventional OFDM system, which maps a symbol
exclusively on one
dedicated subcarrier. $\m{G}_d''$, on the other hand, coincides with
$\m{G}''$ in spreading a symbol almost uniformly over all subcarriers, thus
appearing similar to a single-carrier based system, which spreads it exactly
uniformly over the available bandwidth.
%
%
%

\subsection{Optimization of Pilot Dependent Term}\label{sec:opt_pilot_genmtx}

\begin{figure}[!htb]
 \centering
  \scalebox{0.5}{
%
\begin{tikzpicture}

\begin{axis}[%
width=2.195in,
height=3.566in,
at={(0.368in,0.481in)},
scale only axis,
point meta min=0,
point meta max=1,
axis on top,
xmin=0.5,
xmax=4.5,
xtick={1,2,3,4},
xticklabels={{0},{1},{2},{3}},
xlabel style={font=\color{white!15!black}},
xlabel={\scalebox{1.7}{$\left|\bf{G}_p\right|$}},
y dir=reverse,
ymin=0.5,
ymax=52.5,
axis background/.style={fill=white},
legend style={legend cell align=left, align=left, draw=white!15!black},
colormap={mymap}{[1pt] rgb(0pt)=(1,1,1); rgb(64pt)=(0,0,0)},
colorbar
]
\addplot [forget plot] graphics [xmin=0.5, xmax=4.5, ymin=0.5, ymax=52.5] {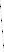};
\end{axis}
\end{tikzpicture}%
}
  \caption{Magnitude of entries of pilot generator matrix $\m{G}_p$ based on
    the setup in Tab.~\ref{tab:setups}.}
  \label{fig:Gp_abs}
\end{figure}

The pilot symbols $\ve{p}$ can usually be freely chosen at design time. Therefore, the pilot
dependent term offers an additional degree of freedom besides the generator matrix $\m{G}_p$, leading to the optimization problem
\begin{align}
  \breve{\m{G}}_{p},\breve{\ve{p}}=\argmin_{\m{G}_p,\ve{p}}\left\{J_p\right\} \hspace{0.1cm}
  \mathrm{s.t.} \hspace{0.1cm} 		\m{F}_N^{-1}\m{B}\m{G}_p
  = \begin{bmatrix} \m{\Xi} \\ \m{0} \end{bmatrix} \,\land\, \left|p[i]\right|=1, \label{equ:pil008}
\end{align}
with $J_p$ denoting an appropriate cost function, $\m{\Xi}~\in~\mathbb{R}^{N \times (N-N_u)}$ again an arbitrary
matrix, and $\left|p[i]\right|=1$ with $i=0,\dots, N_p-1$ a constant energy
constraint on each individual pilot symbol.
In fact, this optimization problem shows strong similarities with the one for finding
generator matrices of a systematically encoded
UW-OFDM system \cite{Huemer14}. Consequently, let us choose the approach
\begin{equation}
\m{G}_p=\m{P}_p\begin{bmatrix}\m{I} \\ \m{T}_p \end{bmatrix},\label{equ:pil009}
\end{equation}
whereas $\m{I}$ in combination with a permutation matrix
$\m{P}_p\in\{0,1\}^{(N_d+N_r+N_p)\times(N_d+N_r+N_p)}$ places the pilot symbols
at the corresponding subcarrier positions, and $\m{T}_p\in\mathbb{C}^{(N_d+N_r)
  \times N_p}$ again ensures the zero-word constraint. Although
$N_r$ subcarriers (assuming $N_r=N_u$) would be sufficient to fulfill the zero-word
constraint from a mathematical point of view (compare with its counterpart
$\m{T}_d$ from Sec.~\ref{sec:opt_data_term}), $\m{T}_p$ spreads the required
redundancy over $(N_d+N_r)$ subcarriers instead, thus providing additional $N_d$ degrees of freedom in the design of the generator matrix.
The matrix $\m{T}_p$ is calculated as $\m{T}_p=-\m{M}''^{\dagger}_{22}\m{M}''_{21}$ with the submatrices
$\m{M}''_{21}\in\mathbb{C}^{N_u \times N_p}$ and 
$\m{M}''_{22}\in\mathbb{C}^{N_u \times (N_d+N_r)}$, $N_r=N_u$, derived from
$\m{F}_N^{-1}\m{B}\m{P}_p=\left[\begin{smallmatrix}\m{M}''_{11} &
    \m{M}''_{12}\\ \m{M}''_{21} &
    \m{M}''_{22} \end{smallmatrix}\right]$. Fig.~\ref{fig:Gp_abs} depicts an
exemplary generator matrix $\m{G}_p$. As intended, $\m{G}_p$ fulfills its purpose and places the pilot symbols on the designated
subcarriers. However, $\m{G}_p$ has to ensure the zero-word
constraint at the same time, which is only possible by also spreading (minor)
portions of $\ve{p}$ on other subcarriers (as already mentioned on at least
$N_r=N_u$). To keep spreading at a minimum as
there is no other purpose beyond fulfilling the zero-word constraint,
a reasonable cost function $J_p$ in \eqref{equ:pil008} is one that delivers
$\m{G}_p$ and $\ve{p}$ which induce minimum energy\footnote{In this context,
  spreading redundancy by $\m{T}_p$ on $(N_d+N_r)$ instead of $N_r$ subcarriers helps
  reducing the required energy $J_p$.}.
The energy $E_{p}$ induced by the pilots is given by the cost function\footnote{The
  energy is scaled by $N$, originating from the DFT, to easier link the resulting energies with the number of
  pilots $N_p$ in the subsequent paragraphs.}
\begin{align}
J_{p} &= E_{p}N=\ve{p}^H\m{G}_p^H\m{G}_p\ve{p}.\label{equ:pil029}
\end{align}
The two different factors influencing the resulting energy of the pilots are
the positions of the pilot subcarriers determined by $\m{P}_p$ within $\m{G}_p$, and the values
of the pilot symbols $\ve{p}$.
Since it has been shown in \cite{Cai04} that a uniform distribution of the
pilot subcarriers over the available bandwidth is beneficial, we fix the
position of the pilots accordingly. Hence, the positions are not available as a degree of freedom to minimize \eqref{equ:pil029}. In other words, the
design of $\m{G}_p$ is fixed and there is no
possibility left for optimization of the generator matrix,
leaving the values of the pilot symbols $\ve{p}$ as only optimization
parameters. The original optimization problem in
\eqref{equ:pil008} reduces then to
\begin{equation}
  \breve{\ve{p}}=\argmin_{\ve{p}}\left\{J_{p}\right\} \,\,\land\,\, \left|p[i]\right|=1. \label{}
\end{equation}
The problem can be solved by an exhaustive search, whereas a pilot symbol $p[i]$ is drawn from an alphabet
\begin{equation}
\mathcal{A}=\left\{e^{j\frac{2\pi \kappa}{|\mathcal{A}|}}\right\} \hspace{1cm}\kappa=0\dots |\mathcal{A}|-1\label{equ:pil030}
\end{equation}
with cardinality $|\mathcal{A}|$, resulting in $|\mathcal{A}|^{N_p}$
different combinations in case of $N_p$ pilot subcarriers. Tab.~\ref{tab:min_energy_pilots} shows the corresponding minimum values for
$E_{p}$ as a function of the
cardinality of the pilot symbol alphabet, evaluated for a generator matrix $\m{G}_p$ constructed
according to the setup parameters in Tab.~\ref{tab:setups} and depicted in
Fig.~\ref{fig:Gp_abs}. For instance, the
global minimum energy of 5.1783 in Tab.~\ref{tab:min_energy_pilots} follows from
evaluating \eqref{equ:pil029} with the pilot symbols $p[i]$ drawn from
\eqref{equ:pil030} with $\kappa=k[i]$ and $k[i]$ denoting the $i$th element of the vector $\ve{k}=[17,14,3,0]$.
\begin{table}[!htb]
  \caption{Minimum pilot induced energy $E_p$ for an UW-OFDM system with
  $\mathcal{I}_p=\{7,21,43,57\}$ as pilot subcarrier positions.}
\begin{center}
\begin{tabular}{l|rrrrr} \hline
$|\mathcal{A}|$ & 2 & 4 & 6 & 10 & 20\\ \hline\hline
$N\cdot E_p$ & 5.4633 & 5.2423 & 5.1969 & 5.1864 & 5.1783 \\ \hline
\end{tabular}
\label{tab:min_energy_pilots}
\end{center}
\end{table}
We notice from Tab.~\ref{tab:min_energy_pilots} that the performance does not
significantly vary with the cardinality $|\mathcal{A}|$. We conclude that the granularity of
the alphabet and thus the optimization effort for the exhaustive search can be kept at a minimum without facing any significant performance loss.

\section{Estimation of Common Phase Error and Carrier Frequency Offset}\label{sec:cfo_phase_rotation}
A well-known critical issue for OFDM systems is the presence of a CFO. A CFO is the result of an oscillator mismatch between transmitter
and receiver, or of a Doppler effect due to a movement of at least one of the
two. Due to the high sensitivity of OFDM, an accurate estimation and
compensation of the CFO is essential. This section will extend the UW-OFDM model in \eqref{equ:pil019} by
CFO effects and based on that analyze an estimation algorithm for an CFO induced phase
offset. The algorithm is intended to be utilized within the tracking phase
\cite{Classen94} of an overall CFO estimation and compensation task and shall
take over a fine CFO estimation on an OFDM symbol by symbol basis. Due to an
aquisition phase normally preceding the tracking, a CFO can safely be assumed to be
in the range of a fraction of the subcarrier
spacing. In practice, the remaining CFO will be less than
  10\,\% when set in relation to the subcarrier spacing \cite{Classen94}. The 
CFO effects in this section are detailed up to the level necessary for
developing the estimation algorithm in Sec.~\ref{sec:carrier_phase_est}. The
interested reader is therefore referred to App.~\ref{sec:app_a} for details on the
definitions and derivations.


\subsection{Receiver Model}

Assuming a relative\footnote{A relative
  carrier frequency offset $\epsilon$ is defined as
  $\epsilon=\tfrac{f_\cfo}{\Delta_f}$, whereas $f_\cfo$ denotes an absolute
  carrier frequency offset and $\Delta_f$ the absolute subcarrier spacing of the system
  in consideration.} carrier frequency offset $\epsilon$,
%
the receive signal of the $l$th OFDM symbol in the frequency domain incorporating
the CFO effects can be modelled by 
\begin{equation}
\Yfdownl \approx \lambdafl\mf{H}\m{G}_p\ve{p} + \lambdafl\mf{H}\m{G}_d\datal +
\lambdafl\mf{H}\m{B}^T\vef{x}_u +\noisef,\label{equ:cfo_ang006}
\end{equation}
with a minor approximation error that will be investigated in detail in Sec.~\ref{sec:approx_err_rx_model}.
A CFO in the frequency domain modelled by $\lambdafl$ causes three effects on a
subcarrier symbol, which can be explained by the decomposition 
\begin{equation}
\lambdafl=\e^{j\varphil}\mf{\Lambda}_\text{stat}.\label{equ:lambdafl}
\end{equation}
First, the $l$th OFDM symbol is \emph{rotated} by an accumulated phase offset
$\varphil$, which is also referred to as CPE in the literature.
The other two CFO effects are incorporated in the \underline{stat}ic matrix
$\mf{\Lambda}_\text{stat}\in\mathbb{C}^{(N-N_z)\times (N-N_z)}$, which is on
the one hand (in the considered CFO range of $\epsilon\leq 0.1$) a negligible scaling of the subcarrier
symbols represented by the main diagonal entries of $\mf{\Lambda}_\text{stat}$
(e.g., $0.98$ for $\epsilon = 0.1$),
and on the other hand an ICI induced disturbance given by the off-diagonal entries. 

A full compensation of the CFO induced effects would require an inversion of
$\lambdafl$. However, the most dominant out of the three CFO effects is given by $\varphil$
\cite{Classen94}. Hence, an estimation and compensation of $\varphil$ by
derotating with its estimate $\varphilest$ --- also known as common phase
synchronization --- is a sufficient countermeasure
against CFO induced disturbances \cite{Classen94}. The estimation of $\varphil$
will be tackled in Sec.~\ref{sec:carrier_phase_est}.

\subsection{Evaluation of Approximation Error in Receiver Model}\label{sec:approx_err_rx_model}

The receiver model in \eqref{equ:cfo_ang006} is only an approximation, as it
neglects the ICI induced disturbances resulting from those UW spectrum parts
that overlay otherwise unused zero subcarriers. As detailed in the Appendix in \eqref{equ:cfo_rx019},
there would be actually an additional fifth term
$\mf{\Lambda}_{zn}^{(l)}\mf{H}_z\vef{x}_{u,z}$ in \eqref{equ:cfo_ang006}. This term represents that portion of the UW in frequency domain, which overlays potential \underline{z}ero subcarriers, i.e.,
$\vef{x}_{u,z}$, and is then spread on \underline{n}on-zero subcarriers due to
ICI effects modelled within $\mf{\Lambda}_{zn}^{(l)}$\footnote{Note that $\mf{\Lambda}_{zn}^{(l)}$ can be
  decomposed into the same CFO effects as $\lambdafl$ in \eqref{equ:lambdafl}.}. The following part will investigate the approximation error in
\eqref{equ:cfo_ang006} in detail and
confirm that this error is negligible.

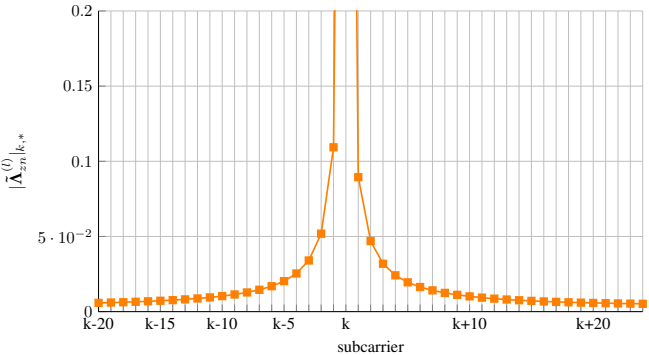
\begin{figure}[!tbh]
\centering
\scalebox{0.63}{
%
\definecolor{mycolor1}{rgb}{1.00000,0.50196,0.00000}%
\begin{tikzpicture}

\begin{axis}[%
width=4.521in,
height=2.5in,
at={(0.758in,0.481in)},
scale only axis,
xmin=10,
xmax=54,
xtick={1,2,3,4,5,6,7,8,9,10,11,12,13,14,15,16,17,18,19,20,21,22,23,24,25,26,27,28,29,30,31,32,33,34,35,36,37,38,39,40,41,42,43,44,45,46,47,48,49,50,51,52,53,54,55,56,57,58,59,60,61,62,63,64},
xticklabels={{},{},{},{},{k-25},{},{},{},{},{k-20},{},{},{},{},{k-15},{},{},{},{},{k-10},{},{},{},{},{k-5},{},{},{},{},{k},{},{},{},{},{},{},{},{},{},{k+10},{},{},{},{},{},{},{},{},{},{k+20},{},{},{},{},{},{},{},{},{},{k+30},{},{},{},{}},
xlabel={subcarrier},
ymin=0,
ymax=0.2,
ylabel={$|\mf{\Lambda}_{zn}^{(l)}|_{k,*}$},
axis background/.style={fill=white},
axis x line*=bottom,
axis y line*=left,
xmajorgrids,
ymajorgrids,
legend style={legend cell align=left, align=left, draw=white!15!black}
]
\addplot [color=mycolor1, line width=1.0pt, mark=square*, mark options={solid, fill=mycolor1, mycolor1}]
  table[row sep=crcr]{%
1	0.00488483809512798\\
2	0.00492785525189881\\
3	0.00498374974002981\\
4	0.00505324011134781\\
5	0.0051372468301943\\
6	0.005236923366738\\
7	0.00535369715258485\\
8	0.00548932325349463\\
9	0.00564595476468479\\
10	0.00582623559296555\\
11	0.00603342372843327\\
12	0.00627155676448345\\
13	0.0065456770209987\\
14	0.00686214238333677\\
15	0.00722906301240151\\
16	0.00765692721182876\\
17	0.00815951900307656\\
18	0.0087552989167419\\
19	0.00946954540047473\\
20	0.0103377945806892\\
21	0.011411599486966\\
22	0.012768663894228\\
23	0.014531788737818\\
24	0.0169070866344735\\
25	0.0202689903121892\\
26	0.0253760439507556\\
27	0.0340331602484001\\
28	0.0518452168529232\\
29	0.109327964843998\\
30	0.983635593312341\\
31	0.0894645256425416\\
32	0.0469226595809505\\
33	0.0318528413839702\\
34	0.0241537432443592\\
35	0.0194898389183543\\
36	0.0163686167976719\\
37	0.0141384583126586\\
38	0.0124695696693621\\
39	0.0111771566831552\\
40	0.0101496445448359\\
41	0.0093156979910589\\
42	0.00862759539045739\\
43	0.00805221804078077\\
44	0.00756587410766524\\
45	0.00715117959179099\\
46	0.00679510200325959\\
47	0.00648769082459558\\
48	0.00622122920204684\\
49	0.00598965254284865\\
50	0.0057881411276427\\
51	0.00561282907770114\\
52	0.00546059290205991\\
53	0.00532889560065838\\
54	0.00521567029200878\\
55	0.0051192324644801\\
56	0.00503821331645675\\
57	0.00497150890465921\\
58	0.00491824135930229\\
59	0.00487772949691609\\
60	0.00484946692444457\\
61	0.00483310628445427\\
62	0.00482844870945122\\
63	0.00483543788092095\\
64	0.00485415835972975\\
};

\end{axis}
\end{tikzpicture}%
}
\caption{Magnitude of ICI effects experienced by the
  $k$th subcarrier from its neighboring subcarriers, with $\epsilon=0.1$ and $|\mf{\Lambda}_{zn}^{(l)}|_{k,k}=0.98$.}
\label{fig:ici_magnitude}
\end{figure}

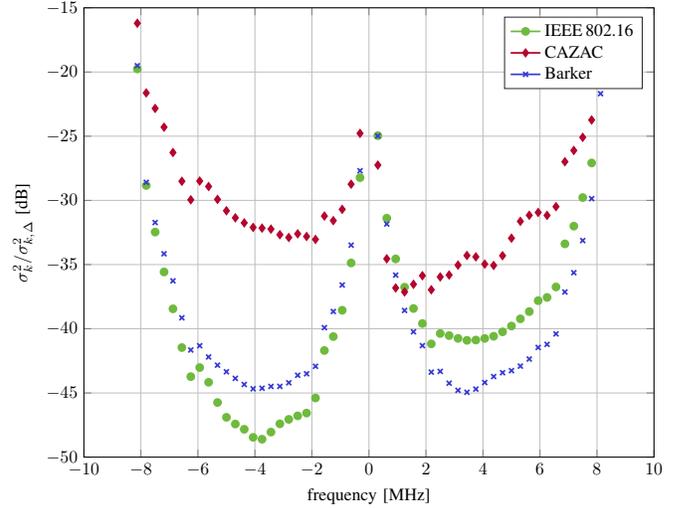
\begin{figure}[!tbh]
\centering
\scalebox{0.66}{
%
\definecolor{mycolor1}{rgb}{0.43529,0.72941,0.23922}%
\definecolor{mycolor2}{rgb}{0.70588,0.01961,0.18039}%
\definecolor{mycolor3}{rgb}{0.22353,0.23922,0.83137}%
\begin{tikzpicture}

\begin{axis}[%
width=4.521in,
height=3.566in,
at={(0.758in,0.481in)},
scale only axis,
xmin=-10,
xmax=10,
xlabel={frequency [MHz]},
ymin=-50,
ymax=-15,
ylabel={$\sigma^2_{k}/\sigma^2_{k,\Delta}$ [dB]},
axis background/.style={fill=white},
xmajorgrids,
ymajorgrids,
legend style={legend cell align=left, align=left, draw=white!15!black}
]
\addplot [color=mycolor1, line width=1.0pt, draw=none, mark=*, mark options={solid, fill=mycolor1, mycolor1}]
  table[row sep=crcr]{%
-8.125	-19.7606436738237\\
-7.8125	-28.8384893752994\\
-7.5	-32.467292286526\\
-7.1875	-35.5747298191183\\
-6.875	-38.4488932648259\\
-6.5625	-41.4654225961306\\
-6.25	-43.7289403388955\\
-5.9375	-43.0227497372125\\
-5.625	-44.1676549316783\\
-5.3125	-45.7446440493892\\
-5	-46.8986170988331\\
-4.6875	-47.4229151314281\\
-4.375	-47.8284924770987\\
-4.0625	-48.4586163736389\\
-3.75	-48.603620465363\\
-3.4375	-48.0495369465451\\
-3.125	-47.4038235415723\\
-2.8125	-47.0572046046056\\
-2.5	-46.7838753798242\\
-2.1875	-46.5674276892762\\
-1.875	-45.3912685915499\\
-1.5625	-41.6933357599873\\
-1.25	-40.6087943341754\\
-0.9375	-38.5593431079338\\
-0.625	-34.8725306030857\\
-0.3125	-28.2291534404403\\
0.3125	-24.9597970934971\\
0.625	-31.3864964066927\\
0.9375	-34.5587362349144\\
1.25	-36.7623889377428\\
1.5625	-38.4204296143134\\
1.875	-39.5933636754078\\
2.1875	-41.1754082295429\\
2.5	-40.3680419990545\\
2.8125	-40.5313036558398\\
3.125	-40.7499798453451\\
3.4375	-40.9019746640966\\
3.75	-40.8784329656678\\
4.0625	-40.753000288567\\
4.375	-40.5832124578581\\
4.6875	-40.2396990590316\\
5	-39.7875919993268\\
5.3125	-39.226007957445\\
5.625	-38.6619411884684\\
5.9375	-37.8135960798753\\
6.25	-37.5606798934428\\
6.5625	-36.745077910225\\
6.875	-33.3861753705517\\
7.1875	-32.0055461672197\\
7.5	-29.7796349164604\\
7.8125	-27.084676580339\\
8.125	-19.5325154480626\\
};
\addlegendentry{IEEE\,802.16}

\addplot [color=mycolor2, line width=1.0pt, draw=none, mark=diamond*, mark options={solid, fill=mycolor2, mycolor2}]
  table[row sep=crcr]{%
-8.125	-16.2025418572796\\
-7.8125	-21.6286326031699\\
-7.5	-22.8365068467416\\
-7.1875	-24.2986636200727\\
-6.875	-26.2723788243116\\
-6.5625	-28.5116090176004\\
-6.25	-29.9555250687743\\
-5.9375	-28.4920927817498\\
-5.625	-28.9181880203135\\
-5.3125	-29.9227275300653\\
-5	-30.8076122377997\\
-4.6875	-31.357431911393\\
-4.375	-31.7517436671999\\
-4.0625	-32.1053167167022\\
-3.75	-32.1606470678157\\
-3.4375	-32.2392414101599\\
-3.125	-32.6735265023034\\
-2.8125	-32.8904099486934\\
-2.5	-32.6040850289661\\
-2.1875	-32.8119458901524\\
-1.875	-33.0394011339645\\
-1.5625	-31.2202439186877\\
-1.25	-31.5779402346816\\
-0.9375	-30.7000165867726\\
-0.625	-28.740378447765\\
-0.3125	-24.780491468296\\
0.3125	-27.2517066583218\\
0.625	-34.5508698322384\\
0.9375	-36.8201526310528\\
1.25	-37.1323265259515\\
1.5625	-36.5374012478979\\
1.875	-35.8668180057073\\
2.1875	-36.9620258141966\\
2.5	-35.9631801416956\\
2.8125	-35.8132124264018\\
3.125	-35.0426777878719\\
3.4375	-34.2938579102479\\
3.75	-34.3989834340378\\
4.0625	-34.964306831385\\
4.375	-35.0679018493489\\
4.6875	-34.3076165416183\\
5	-32.9448730970035\\
5.3125	-31.6322822063708\\
5.625	-31.1593258811378\\
5.9375	-30.9402679846936\\
6.25	-31.1684727503129\\
6.5625	-30.488394785343\\
6.875	-26.9889295620726\\
7.1875	-26.111598145451\\
7.5	-25.0889017062627\\
7.8125	-23.7360756895738\\
8.125	-17.7825530597337\\
};
\addlegendentry{CAZAC}

\addplot [color=mycolor3, line width=1.0pt, draw=none, mark=x, mark options={solid, fill=mycolor3, mycolor3}]
  table[row sep=crcr]{%
-8.125	-19.5104176707463\\
-7.8125	-28.5924916642133\\
-7.5	-31.7278207415499\\
-7.1875	-34.1599553593459\\
-6.875	-36.268264616868\\
-6.5625	-39.1478780330288\\
-6.25	-41.6546870154067\\
-5.9375	-41.3219843820367\\
-5.625	-42.1975668608236\\
-5.3125	-42.8371550624427\\
-5	-43.3512142177577\\
-4.6875	-43.8646588512627\\
-4.375	-44.3396190500395\\
-4.0625	-44.6778330897018\\
-3.75	-44.6319794710915\\
-3.4375	-44.4933994376555\\
-3.125	-44.4888208056216\\
-2.8125	-44.2076584091082\\
-2.5	-43.6225407855683\\
-2.1875	-43.5063107131011\\
-1.875	-42.9224850096983\\
-1.5625	-39.9133042448555\\
-1.25	-38.6551392374992\\
-0.9375	-36.5937131693937\\
-0.625	-33.4874464054224\\
-0.3125	-27.6900164353141\\
0.3125	-24.9833622889147\\
0.625	-31.841306516469\\
0.9375	-35.8215417425483\\
1.25	-38.5739468349564\\
1.5625	-40.2342248487956\\
1.875	-41.3132820849446\\
2.1875	-43.377899541319\\
2.5	-43.3206464403712\\
2.8125	-44.2392066466387\\
3.125	-44.7972315301751\\
3.4375	-44.9486621453784\\
3.75	-44.6981543447945\\
4.0625	-44.1895257101991\\
4.375	-43.7242395074545\\
4.6875	-43.424613074421\\
5	-43.2636776394189\\
5.3125	-42.9092949368348\\
5.625	-42.3577297190862\\
5.9375	-41.4554852463772\\
6.25	-41.2154240772588\\
6.5625	-40.3996700195962\\
6.875	-37.1351613128501\\
7.1875	-35.6318604961985\\
7.5	-33.1310800573658\\
7.8125	-29.8657597777091\\
8.125	-21.6877072753295\\
};
\addlegendentry{Barker}

\end{axis}
\end{tikzpicture}%
}
\caption{Evaluation of approximation error in \eqref{equ:cfo_ang006} due to neglecting ICI
  induced disturbances from UWs on non-zero subcarriers. Investigated scenario: $\m{G}'_d$, $\m{G}_p$, $\mf{H}_p=\m{I}$, $\epsilon=0.1$ and $\varnoisef=0$.}
\label{fig:err_lin_model_approx}
\end{figure}

Normally, zero subcarriers follow a dedicated
purpose like shaping the spectral mask. Hence, a UW might be chosen and designed already beforehand
to explicitly maintain these properties and thus justify our
approximation. However, even if this is not possible, one can still simply
subtract the corresponding offset, given the availability of an estimate of $\mf{\Lambda}_{zn}^{(l)}$ and the
channel frequency response $\mf{H}_z$ on the \underline{z}ero subcarriers. The question
remains though, if the effort for estimation and subtraction even translates to
any noticeable performance gain, which is investigated next. For that, we
elaborate on the error introduced by ICIs in principle in a first step (Fig.~\ref{fig:ici_magnitude}), and
complement this elaboration with practical examples in a second step (Fig.~\ref{fig:err_lin_model_approx}).

Overall, the entries of $\mf{\Lambda}_{zn}^{(l)}$ incorporate the potential impact of ICIs. 
Fig.~\ref{fig:ici_magnitude} plots the magnitude of the entries of the $k$th row for
$\epsilon=0.1$, which determine the influence of the neighbors on the $k$th
subcarrier due to ICI.
Since the magnitudes of the ICI coefficients monotonically
increase in $\epsilon$, this denotes the worst-case-scenario for ICI effects
seen in the considered
tracking phase. Obviously,
the impact of ICI declines rapidly as a function of the distance, from
0.1 for the first to already 0.03 for the third neighboring subcarrier. From a
practical point of view, this means that independent of the actual UW load on
the subcarriers, only the close neighboring subcarriers are
likely to cause any relevant ICI induced disturbances. Consequently, these
observations already strongly indicate that the approximation error shall
always be rather small. 
In order to complete investigations, Fig.~\ref{fig:err_lin_model_approx} examines the resulting ICI effects
evaluated for actual UWs, namely a CAZAC \cite{Popovic1992}, a Barker
\cite{Golomb1965} and the UW sequence from IEEE\,802.16 \cite{IEEE80216}. It depicts for each utilized (i.e., non-zero) subcarrier
the ratio between the average power of the approximated signal in
\eqref{equ:cfo_ang006} with
\begin{equation}
\sigma^2_{k} = \E{\tilde{y}_\downs[k]\tilde{y}_\downs[k]^H},
\end{equation}
and the average power of the approximation error
\begin{equation}
\sigma^2_{k,\Delta} = \E{\elem{\left(\mf{\Lambda}_{zn}\mf{H}_z\vef{x}_{u,z}\right)\left(\mf{\Lambda}_{zn}\mf{H}_z\vef{x}_{u,z}\right)^H}_{k,k}},
\end{equation}
which corresponds to the ICI induced disturbances originating from the UW load on zero
subcarriers. It turns out that the impact of an ICI
induced UW error is in a negligible range, at most at the edges of the
utilized bandwidth it might be in a noticeable range. In conclusion, the
approximation error is non-dominant in practical scenarios and therefore
justifies the simplification in \eqref{equ:cfo_ang006}. Hence, in the following investigations the approximate model in \eqref{equ:cfo_ang006} can be used.

\subsection{Estimation of CPE $\varphil$}\label{sec:carrier_phase_est}

There are various approaches for estimating the CPE $\varphil$ in
OFDM systems, e.g., based on data symbols \cite{Kai05,Ren08} known as decision directed schemes or
cyclic prefix based methods \cite{Keller01,Lashkarian00}. Another common way is the
utilization of pilot symbols $\ve{p}$ in the frequency domain \cite{Classen94}
to obtain the estimate
\begin{equation}
\varphilest = \text{arg}\left(\ve{p}^H\m{W}_p\pestl\right).\label{equ:cfo_ang005}
\end{equation}
The estimate $\pestl$ is attained from the
$l$th received OFDM symbol in \eqref{equ:cfo_ang006} and $\m{W}_p=\diag{\ve{w}_p}$ is a diagonal weighting
matrix with $\m{w}_p\in{\mathbb{R}^0_+}^{(N_p\times 1)}$ to rate the pilots regarding their estimation quality (e.g., based on the inverse
main diagonal of an error covariance matrix). In order to evaluate the applicability of this concept to UW-OFDM as well, let us
investigate the CFO effects on the pilot symbols in more detail. The
general definition of a received OFDM symbol introduced in
\eqref{equ:cfo_ang006} serves as a starting point to obtain $\pestl$. We apply
a linear estimator $\m{E}_p'=\m{E}_p\mf{H}^{-1}$ consisting of two stages. 
The first stage inverts the channel, and the second stage $\m{E}_p=\begin{bmatrix}\m{I} & \m{0} \end{bmatrix}\m{P}_p^T$
extracts the pilots from the frequency domain vector, with the permutation matrix
$\m{P}_p$ from \eqref{equ:pil009}. We are aware that $\m{E}_p$ is
only suboptimum in terms of estimation performance, as it is not capable of exploiting the portion of the
pilots spread on the non-pilot subcarriers due to $\m{G}_p\ve{p}$. This for
instance would be possible with a best linear unbiased estimator (BLUE)
$\m{E}_p=\m{G}_p^\dagger$ \cite{Kay93}. Nevertheless,
this suboptimum approach does e.g., not require the subtraction of the data and UW dependent offset term
in \eqref{equ:cfo_ang006} to transform an affine into a linear model
\cite{Kay93}, which therefore simplifies the estimation task greatly. Applying $\m{E}_p'$ leads to
\begin{align}
 \pestl&=\m{E}_p\mf{H}^{-1}\Yfdownl\label{equ:cfo_ang011}\\
  &=  \m{E}_p\mf{H}^{-1}\lambdafl\mf{H}\m{G}_p\ve{p}
  + \m{E}_p\mf{H}^{-1}\lambdafl\mf{H}\m{G}_d\datal \nonumber\\
&\quad  +\m{E}_p\mf{H}^{-1}\lambdafl\mf{H}\m{B}^T\vef{x}_u +
  \m{E}_p\mf{H}^{-1}\m{B}^T\noisef\label{equ:cfo_ang012}\\
    &= \m{E}_p\lambdafHl\left(\m{G}_p\ve{p} + \m{B}^T\vef{x}_u\right) + \dataicil + \noisefpil.\label{equ:cfo_ang013}
\end{align}
Pilot and data symbols are orthogonal in
frequency domain, but $\lambdafl$ introduces \underline{i}nter\underline{c}arrier \underline{i}nterference resulting in
\begin{equation}
\dataicil=\m{E}_p\mf{H}^{-1}\lambdafl\mf{H}\m{G}_d\datal\label{equ:cfo_ang013a}, 
\end{equation}
with
$\dataicil\in\mathbb{C}^{N_p\times 1}$. The vector
$\noisefpil\in\mathbb{C}^{N_p\times 1}$ represents additive noise according to
\begin{equation}
\noisefpil=\m{E}_p\mf{H}^{-1}\m{B}^T\noisef.
\end{equation}
Since $\lambdafl$ has entries off
the main diagonal and is only
approximately a diagonal matrix, multiplying with $\mf{H}^{-1}$ cannot fully
equalize the channel $\mf{H}$, which is incorporated in
\begin{equation}
  \lambdafHl=\mf{H}^{-1}\lambdafl\mf{H}.
\end{equation}
For an easier analysis, a \emph{single}
pilot symbol is considered in the following, which is given as 
\begin{align}
\kElempestl&=\mel{\m{E}_p}_{k,*}\lambdafHl\left(\m{G}_p\ve{p} + \m{B}^T\vef{x}_u\right) + \kElemdataicil +\kElemNoisePil\nonumber\\
&=
\mel{\m{E}_p}_{k,*}\lambdafHl\left(\mel{\m{G}_p}_{*,k}\kElemp+
\m{B}^T\vef{x}_u\right)
\mel{\m{E}_p}_{k,*}\lambdafHl\nonumber\\
&\quad\times\sum_{m=0,m\neq
k}^{N_p-1}\mel{\m{G}_p}_{*,m}\mElemp 
 + \kElemdataicil +\kElemNoisePil\nonumber\\
&= \ve{e}^T_{k}\lambdafHl\left(\ve{g}_{k}\kElemp+ \m{B}^T\vef{x}_u\right)\nonumber\\
&\quad+
\ve{e}^T_{k}\lambdafHl\sum_{m=0,m\neq k}^{N_p-1}\ve{g}_{m}\mElemp +
    \kElemdataicil +\kElemNoisePil\nonumber\\
&= \ve{e}^T_{k}\lambdafHl\left(\ve{g}_{k}\kElemp+ \m{B}^T\vef{x}_u\right)
+ \kElempicil + \kElemdataicil +\kElemNoisePil,\label{equ:cfo_ang016a}
\end{align}
with $\ve{e}^T_{k}=\mel{\m{E}_p}_{k,*}$, $\ve{g}_{k}=\mel{\m{G}_p}_{*,k}$ and $\ve{g}_{m}=\mel{\m{G}_p}_{*,m}$. Consequently, there are four terms influencing $\kElempestl$. The additive
noise term $\kElemNoisePil$ should not require any further explanation. The term $\kElemdataicil$ models the ICI driven by data symbols and degrades
estimation quality in a similar manner as additive noise. The term
\begin{equation}
\kElempicil=\ve{e}^T_{k}\lambdafHl\sum_{m=0,m\neq k}^{N_p-1}\ve{g}_{m}\mElemp\label{equ:cfo_ang042}
\end{equation}
comprises the ICI induced by the
other pilot symbols. Since the pilot symbols are constant, this term leads to a
constant estimation error (assuming a fixed CFO $\epsilon$). In any way, the
influence of this term is rather limited. This is due to a usually uniform distribution of the pilot symbols
over the available spectrum to optimally support system parameter estimation
tasks \cite{Cai04}, resulting in a distance of several
subcarriers among them, cf.~Fig.~\ref{fig:Gp_abs}. The first term in \eqref{equ:cfo_ang016a} comprises the actual pilot symbol, an additive offset
caused by the UW and a part causing the CPE to be estimated. 

Let the last three terms in
\eqref{equ:cfo_ang016a} form a new additive
noise term $\kElemNNoisePil=\kElempicil + \kElemdataicil +\kElemNoisePil$, then
the estimation of $\varphil$ reads
\begin{align}
  \varphilest&= \text{arg}\left(\ve{p}^H\m{W}_p\pestl\right)\label{equ:cfo_ang017}\\
  &=\text{arg}\left(\sum_{k=0}^{N_p-1}\kElemp^H\kElemwp\kElempestl\right)\label{equ:cfo_ang018}\\
&=\text{arg}\Biggl(\sum_{k=0}^{N_p-1}\ve{e}^T_{k}\lambdafHl\left(\ve{g}_{k}\kElemwp|\kElemp|^2
  + \m{B}^T\vef{x}_u\kElemwp\kElemp^H\right)\nonumber\\
&\quad  + \kElemNNoisePil\kElemwp\kElemp^H\Biggr)\label{equ:cfo_ang019}\\
&=\text{arg}\Biggl(\underbrace{\sum_{k=0}^{N_p-1}\e^{j\varphil}\ve{e}^T_{k}\mf{\Lambda}_{h,\text{stat}}\kElemwp\left(\ve{g}_{k}|\kElemp|^2
  + \m{B}^T\vef{x}_u\kElemp^H\right)}_{\text{pilot term}}\nonumber\\
&\quad  +
  \underbrace{\sum_{k=0}^{N_p-1}\kElemNNoisePil\kElemwp\kElemp^H}_{\text{noise
      term}}\Biggr),\label{equ:cfo_ang020}
\end{align}
with $\mf{\Lambda}_{h,\text{stat}}=\mf{H}^{-1}\mf{\Lambda}_\text{stat}\mf{H}$.
The angle $\varphilest$ depends now on two sources, with the first one
denoted as \emph{pilot term} and the second one as \emph{noise term}. Through investigation of the individual terms in
\eqref{equ:cfo_ang020} it can be shown that the noise term has only a minor
impact on $\text{arg}(\cdot)$, i.e., \eqref{equ:cfo_ang020}
almost equals an argument soley taken from the pilot term. This translates to the model
\begin{align}  
\varphilest&=\text{arg}\Biggl(\sum_{k=0}^{N_p-1}\e^{j\varphil}\ve{e}^T_{k}\mf{\Lambda}_{h,\text{stat}}\bigl(\ve{g}_{k}\kElemwp|\kElemp|^2\nonumber\\
 &\quad + \m{B}^T\vef{x}_u\kElemwp\kElemp^H\bigr)\Biggr)
  + \Delta_{l}\label{equ:cfo_ang021}\\
&=\text{arg}\Biggl(\e^{j\varphil}\sum_{k=0}^{N_p-1}\kElemwp\bigl(\ve{e}^T_{k}\mf{\Lambda}_{h,\text{stat}}\ve{g}_{k}|\kElemp|^2\nonumber\\
&\quad  + \ve{e}^T_{k}\mf{\Lambda}_{h,\text{stat}}\m{B}^T\vef{x}_u\kElemp^H\bigr)\Biggr)
  + \Delta_{l},\label{equ:cfo_ang022}
\end{align}
with $\Delta_{l}$ denoting minor additive deviation approximated by
\begin{equation}
\Delta_l\approx f\left(\sum_{k=0}^{N_p-1}\kElemNNoisePil\kElemwp\kElemp^H\right).\label{equ:cfo_ang023}
\end{equation}
We assume $f(\cdot)$ to be a proper function that takes care (in a not further specified way) of
shifting the deviation outside of $\text{arg}(\cdot)$. A detailed specification
of the function $f(\cdot)$ is not of relevance for the subsequent analysis and is therefore disregarded. The expression in
\eqref{equ:cfo_ang022} is still quite complex and requires additional
simplification to allow an intuitive interpretation of the impact factors on
$\varphilest$. Let us therefore introduce the
magnitude and phase representation
\begin{align}
a_{\pil,k}\e^{j\varphi_{\pil,k}}&=\kElemwp\ve{e}^T_{k}\mf{\Lambda}_{h,\text{stat}}\left(\ve{g}_{k}|\kElemp|^2
+ \m{B}^T\vef{x}_u\kElemp^H\right)\label{equ:cfo_scalar001}\\ 
a_{\pil}\e^{j\varphi_{\pil}}&=\sum_{k=0}^{N_p-1}a_{\pil,k}\e^{j\varphi_{\pil,k}},\label{equ:cfo_scalar}
\end{align}
then the estimated CPE $\varphilest$ can finally be written as
\begin{align}
\varphilest
&= \text{arg}\left(\e^{j\varphil}\sum_{k=0}^{N_p-1}a_{\pil,k}\e^{j\varphi_{\pil,k}}\right)
  + \Delta_{l}\label{equ:cfo_ang025}\\
&=
  \text{arg}\left(\e^{j\varphil}a_{\pil}\e^{j\varphi_{\pil}}\right)
  + \Delta_{l}\label{equ:cfo_ang026}\\  
&=\varphil + \varphi_{\pil} + \Delta_{l}.\label{equ:cfo_ang027}
\end{align}
We observe that the estimate $\varphilest$ in \eqref{equ:cfo_ang027} consists of
the true CPE $\varphil$ disturbed by
$\varphi_{\pil}$ and $\Delta_{l}$. While $\Delta_{l}$ occurs in UW-OFDM and
CP-OFDM (but with different values), the phase offset $\varphi_{\pil}$ is only
present in UW-OFDM. 
As observable from \eqref{equ:cfo_scalar001}, this is due to a non-zero UW and an additional scaling and rotating of
the pilot symbol. The latter originates from the fact that
$\m{G}_p$ places a pilot symbol at the dedicated subcarrier position, but also spreads
portions of it over the remaining subcarriers to fulfill the zero-word
constraint of UW-OFDM, cf.~\eqref{equ:pil003}. In combination with $\lambdafl$, the portions of a pilot symbol spread over several subcarriers are
leaked back. This leads to an ICI induced self interference
of the pilot symbols, which has to be accounted for. Therefore, we present an
approach in the following to compensate the induced
phase offset $\varphi_{\pil}$ within the estimate 
$\varphilest$.



\subsection{Compensation of Phase Offset $\varphi_{\pil}$ in $\varphilest$}\label{sec:comp_phase_offset}

The first step towards offset compensation in \eqref{equ:cfo_ang027} is now to obtain an estimate of
$\varphi_\pil$. There are two problems emerging in this context as a
result of the dependence of
$\varphi_\pil$ on $\epsilon$ due to $\mf{\Lambda}_{h,\text{stat}}$. First and
addressed in this section,
a straightforward relationship between $\varphi_\pil$ and $\epsilon$ is
not immediately apparent
from \eqref{equ:cfo_scalar001}--\eqref{equ:cfo_scalar}. Second and addressed
subsequently in
Sec.~\ref{sec:estimation_cfo}, an estimate
$\est{\varphi}_\pil$ requires also an estimate $\est{\epsilon}$, which in
turn has somehow be obtained from $\varphilest$ based on the relationship
between $\varphil$ and $\epsilon$ (see \eqref{equ:varphil} in the Appendix), thus leading to mutual dependencies among $\varphil$,
$\varphi_\pil$ and $\epsilon$.

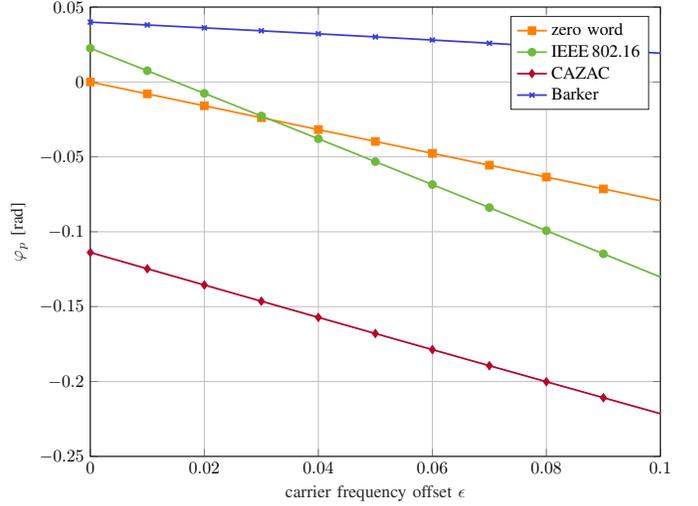
\begin{figure}[!tbh]
\centering
\scalebox{0.66}{
%
\definecolor{mycolor1}{rgb}{1.00000,0.50196,0.00000}%
\definecolor{mycolor2}{rgb}{0.43529,0.72941,0.23922}%
\definecolor{mycolor3}{rgb}{0.70588,0.01961,0.18039}%
\definecolor{mycolor4}{rgb}{0.22353,0.23922,0.83137}%
\begin{tikzpicture}

\begin{axis}[%
width=4.521in,
height=3.566in,
at={(0.758in,0.481in)},
scale only axis,
xmin=0,
xmax=0.1,
xtick={0,0.02,0.04,0.06,0.08,0.1},
xticklabels={{$0$},{$0.02$},{$0.04$},{$0.06$},{$0.08$},{$0.1$}},
xlabel style={font=\color{white!15!black}},
xlabel={carrier frequency offset $\epsilon$},
ymin=-0.25,
ymax=0.05,
ytick={-0.25,-0.2,-0.15,-0.1,-0.05,0,0.05},
yticklabels={{$-0.25$},{$-0.2$},{$-0.15$},{$-0.1$},{$-0.05$},{$0$},{$0.05$}},
ylabel style={font=\color{white!15!black}},
ylabel={$\varphi_{\pil}$ [rad]},
axis background/.style={fill=white},
xmajorgrids,
ymajorgrids,
legend style={legend cell align=left, align=left, draw=white!15!black}
]
\addplot [color=mycolor1, line width=1.0pt, mark=square*, mark options={solid, fill=mycolor1, mycolor1}]
  table[row sep=crcr]{%
1e-06	-7.95373403566486e-07\\
0.010001	-0.0079522859216069\\
0.020001	-0.0158992955220026\\
0.030001	-0.0238418298797222\\
0.040001	-0.0317798912239247\\
0.050001	-0.0397134782810091\\
0.060001	-0.0476425862418923\\
0.070001	-0.0555672067231008\\
0.080001	-0.0634873277217171\\
0.090001	-0.0714029335638696\\
0.100001	-0.0793140048465882\\
};
\addlegendentry{zero word}

\addplot [color=mycolor2, line width=1.0pt, mark=*, mark options={solid, fill=mycolor2, mycolor2}]
  table[row sep=crcr]{%
1e-06	0.0225315273739469\\
0.010001	0.00751130034791406\\
0.020001	-0.00757463756077812\\
0.030001	-0.0227231958804095\\
0.040001	-0.0379311485217903\\
0.050001	-0.0531951338544418\\
0.060001	-0.0685116552332653\\
0.070001	-0.0838770819912543\\
0.080001	-0.0992876509116998\\
0.090001	-0.114739468190215\\
0.100001	-0.130228511893935\\
};
\addlegendentry{IEEE\,802.16}

\addplot [color=mycolor3, line width=1.0pt, mark=diamond*, mark options={solid, fill=mycolor3, mycolor3}]
  table[row sep=crcr]{%
1e-06	-0.113834831873004\\
0.010001	-0.12470787469347\\
0.020001	-0.135555512524623\\
0.030001	-0.146378216645892\\
0.040001	-0.157176406848703\\
0.050001	-0.167950449817126\\
0.060001	-0.178700657524921\\
0.070001	-0.189427285648784\\
0.080001	-0.200130531998009\\
0.090001	-0.210810534960651\\
0.100001	-0.221467371966443\\
};
\addlegendentry{CAZAC}

\addplot [color=mycolor4, line width=1.0pt, mark=x, mark options={solid, fill=mycolor4, mycolor4}]
  table[row sep=crcr]{%
1e-06	0.0399649677834805\\
0.010001	0.0380681725463709\\
0.020001	0.0361349586756795\\
0.030001	0.0341641449387774\\
0.040001	0.0321545115640893\\
0.050001	0.0301047979365555\\
0.060001	0.0280137001876942\\
0.070001	0.0258798686716735\\
0.080001	0.0237019053179493\\
0.090001	0.021478360850823\\
0.100001	0.0192077318655353\\
};
\addlegendentry{Barker}

\end{axis}
\end{tikzpicture}%
}
\caption{UW-OFDM specific phase offset $\varphi_{\pil}$  for
  different UWs
  when estimating the CPE $\varphil$ caused by a CFO. Investigated
  scenario: $\m{G}_p$, 
  $\mf{H}_p=\m{I}$, $\varnoisef=0$ and $\m{W}_p=\m{I}$.}
\label{fig:phi_pil_offset}
\end{figure}

We have confirmed through empirical investigations that
$\varphi_{\pil}$ can be modelled as
\begin{equation}
\varphi_{\pil}=m\epsilon + q,\;\;\;\;\;\;\; m,q\,\in\mathbb{R},\label{equ:cfo_ang037}
\end{equation}
in the relevant CFO range of $\epsilon\leq 0.1$. This behavior is exemplarily
shown in Fig.~\ref{fig:phi_pil_offset} for the pilot generator matrix $\m{G}_p$
visualized in Fig.~\ref{fig:Gp_abs} in case of $\mf{H}_p=\m{I}$, $\m{W}_p=|\mf{H}_p|^2=\m{I}$
and different UWs. Here, $\mf{H}_p\in\mathbb{C}^{N_p\times N_p}$
denotes a diagonal matrix with the channel coefficients corresponding to the
pilot subcarriers on the main diagonal. 
It turns out that for
$\vef{x}_u\neq\ve{0}$, the offset is approximately an affine function of $\epsilon$ for the relevant
CFO range (i.e., $\epsilon\leq 0.1$) for a given setup, which collapses to a
linear function in case of $\vef{x}_u=\ve{0}$. In particular, if
$\epsilon=0$, then the matrix $\mf{\Lambda}_{h,\text{stat}}$ in
\eqref{equ:cfo_scalar001} collapses to an identity matrix for any realization of
$\mf{H}$, resulting in
$\ve{e}^T_{k}\mf{\Lambda}_{h,\text{stat}}\ve{g}_k=1$ and
$\ve{e}^T_{k}\mf{\Lambda}_{h,\text{stat}}\m{B}^T\vef{x}_u=\IpkElemUW$,
with $i_{p,k}$ addressing the $k$th element of the \underline{o}rdered \underline{p}ilot subcarrier
index set $\mathcal{I}_{p,o}=\left(\mathcal{I}_p,<\right)$ defined as
\begin{equation}
\mathcal{I}_{p,o}=\left\{i_{p,0},i_{p,1},\dots i_{p,N_p-1} \right\}.
\end{equation}
In this context,
an estimate follows then as
\begin{align}
  \begin{split}
\est{\varphi}_{\pil|\epsilon=0}&=q\\
&=\arg\left(\sum_{k=0}^{N_p-1}\kElemwp\left(|\kElemp|^2+\IpkElemUW\kElemp^H\right)\right)+\Delta_{l}\\
&\approx\arg\left(\sum_{k=0}^{N_p-1}\kElemwp\left(|\kElemp|^2+\IpkElemUW\kElemp^H\right)\right).\label{equ:cfo_ang040}
\end{split}
\end{align}
Clearly, the constant phase offset $q$ originates from the presence of a non-zero UW
and vanishes in case of $\vef{x}_u=\ve{0}$ and thus $\IpkElemUW=0$. The actual value of $q$ depends on the
UW offset at the specific subcarriers and the pilot symbols themselves. Note
that the affine model in \eqref{equ:cfo_ang037} could easily be transformed into a linear model by the
definition of new pilot symbols $\ve{p}'=\ve{p}+\vef{x}_{u,\mathcal{I}_p}$, where
$\vef{x}_{u,\mathcal{I}_p}$ denotes a vector with elements out of $\vef{x}_u$
at the corresponding pilot subcarrier positions. The estimation of $\varphil$ follows then according to
$\varphilest=\arg\left(\ve{p}'^H\m{W}_p\pestl\right)$.

The variable part $m\epsilon$ within the phase offset model for
$\varphi_{\pil}$ exists independent of the
presence of a zero or non-zero UW. The
slope $m$ is determined by both sources, the UW and the rotation and scaling
of the pilot symbols
\begin{align}
  \begin{split}
m\epsilon&=
\arg\left(\sum_{k=0}^{N_p-1}\ve{e}^T_{k}\mf{\Lambda}_{h,\text{stat}}\ve{g}_{k}\kElemwp|\kElemp|^2\right.\\
  &\left.\quad+
  \ve{e}^T_{k}\mf{\Lambda}_{h,\text{stat}}\m{B}^T\vef{x}_u\kElemwp\kElemp^H\vphantom{\ve{e}^T_{k}}\right)
  -q,\label{equ:cfo_rx041}
  \end{split}
\end{align}
which in turn depend on the channel realization $\mf{H}$, the generator
matrix $\m{G}_p$, the estimator $\m{E}_p$, and the pilot symbols
$\ve{p}$. Regardless of the specific values of these parameters, the
affine model for $\varphi_{\pil}$ as a function of $\epsilon$ holds in the relevant range.

In a real
communication system, the parameters of the affine model for
$\varphi_\pil$ in \eqref{equ:cfo_ang037} are easily determined. Given a certain setup and assuming knowledge of the channel
$\mf{H}$ or an estimate of it (which does not induce an additional effort as it is required for other purposes anyway), the parameters can be
derived by numerically evaluating $\varphi_\pil$ at two different points,
e.g., for $\epsilon=0$ and $\epsilon=0.1$. 
Inserting then \eqref{equ:cfo_ang037}
into \eqref{equ:cfo_ang027} yields
\begin{align}
\varphilest&=\varphil+\varphi_\pil+\Delta_l\label{equ:cfo_ang029}\\
&=\varphil+m\epsilon + q + \Delta_l.\label{equ:cfo_ang030}
\end{align}
Finally, compensating the offset $\varphi_\pil$ by its estimate $\est{\varphi}_\pil$
delivers a new estimate for the CPE
\begin{equation}
\hat{\hat{\varphi}}_l = \varphilest - \hat{\varphi}_\pil
= \varphilest - m\hat{\epsilon} -q\label{equ:cfo_ang031}.
\end{equation}


\subsection{Estimation of CFO $\epsilon$}\label{sec:estimation_cfo}
The presented linearization in \eqref{equ:cfo_ang037} highlights the simple
relationship between the phase offset $\varphi_\pil$  and the CFO
$\epsilon$, thus allowing for an easy compensation. In a real system,
however, $\varphi_\pil$ is not
available and we have to obtain an estimate
\begin{equation}
  \est{\varphi}_\pil = m\hat{\epsilon}+q
\end{equation}
instead, which in turn requires an estimate $\est{\epsilon}$. This will be provided next. The obvious way is exploiting the linear
relationship between $\varphi_l$ and the CFO $\epsilon$ from
\eqref{equ:varphil}. In practice, there is only the estimate
$\varphilest$ avaible which incorporates the unknown
offset $\varphi_{\pil}$ as detailed in \eqref{equ:cfo_ang027}, thus leading to mutual dependencies. Fortunately, the approximation of $\varphi_{\pil}$ as a linear function of
$\epsilon$ resolves these dependencies. Introducing the
prefactor $\frac{2\pi}{N}\epsilon$ to model
$m=N_\pil\frac{2\pi}{N}\epsilon$ with a constant
$N_\pil\in\mathbb{R}$, and using \eqref{equ:varphil} together with
\eqref{equ:cfo_ang030} yields
\begin{align}
  \est{\epsilon}_l &= \epsilon + \Delta_{\epsilon_l}\\
&=  \left(\varphilest-q\right)\frac{N}{2\pi\left( Nl + N_u +\frac{N-1}{2}
  + N_\pil\right)}.\label{equ:cfo_ang032}
\end{align}
The error $\Delta_{\epsilon_l}$ in $\hat{\epsilon}_l$ decreases with increasing
$\varphilest$ due to $Nl$ in the denominator. However, the proposed estimation algorithm $\varphilest= \text{arg}\left(\ve{p}^H\m{W}_p\pestl\right)$ implicitly applies a modulo
operation on $\varphilest$ over the range $[0,2\pi)$,
whereas the estimator in \eqref{equ:cfo_ang032} requires the total angle 
accumulated from the beginning of the burst up to and including OFDM symbol $l$. This restriction
limits the applicability to angles $\varphil$ not exceeding $2\pi$ and thus
also the estimation accuracy. One way to circumvent this restriction is averaging over several estimates $\hat{\epsilon}_l$, gained from
the knowledge that the angle increases linearly with
\begin{align}
\varphi_\Delta &= \varphil - \varphi_{l-1}= \frac{2\pi}{N}\epsilon\left(Nl - N(l-1)\right) = 2\pi\epsilon\label{equ:cfo_ang004}
\end{align}
between two OFDM symbols. Furthermore, the latter approach offers the
advantage of automatically cancelling out the offset $\est{\varphi}_\pil$
within $\varphi_\Delta$.

At this point, let us briefly summarize the presented algorithm for CFO
compensation. The most critical CFO effect is known as CPE, which
can be estimated as shown in Sec.~\ref{sec:carrier_phase_est}. However, in case of UW-OFDM, this estimate
requires a compensation of the incorporated phase offset
$\varphi_p$. Fortunately, this phase offset can easily be compensated by
approximating it as an affine function of $\epsilon$ as detailed in
Sec.~\ref{sec:comp_phase_offset}. This approximation requires in turn an
estimate $\est{\epsilon}$, which can be obtained as shown in Sec.~\ref{sec:estimation_cfo}.

\section{Performance Evaluation}\label{sec:performance}
The preceding analysis has unveiled a few differences in pilot
based CFO estimation between
UW-OFDM and CP-OFDM. The question remains, whether these differences have an impact on the estimation
quality, which is discussed next.

\subsection{Simulation Setup}
Tab.~\ref{tab:setups} summarizes the most important setup parameters of the
UW-OFDM and CP-OFDM systems in consideration. In order to provide a fair comparison with CP-OFDM, the utilized non-systematically
encoded UW-OFDM generator matrices are scaled such that
$\m{G}_d^H\m{G}_d=\alpha\m{I}$ with $\alpha=N'_{d}/N_d$, whereas
$N'_{d}$ denotes the number of data subcarriers of the reference CP-OFDM
system. This ensures that the data induced mean power per non-pilot subcarrier is the same for both
systems; an important aspect, as this defines the severity of the disturbance caused by
$\dataicil$, cf.~\eqref{equ:cfo_ang013}. This is even slightly advantageous for CP-OFDM, since in UW-OFDM the
total mean power per non-pilot subcarrier is even higher due to the spread of
the pilots according to $\m{G}_p\ve{p}$. Within the presented signaling framework, CP-OFDM
is modelled by the generator matrices $\m{G}_{d,\cp}=\m{B}_p\m{I}$ and $\m{G}_{p,\cp}=\m{P}_{p}\begin{bmatrix}\m{I} &
\m{0}^T \end{bmatrix}^T$, assuming appropriately sized identity and zero
matrices. Burstwise transmission of OFDM symbols is applied, for details on the whole
transmission chain (data modulation alphabet, channel encoder, etc.), the
interested reader is referred to \cite{Hofbauer16}. 

Zero additive noise $\noisefpil=\ve{0}$, cf.~\eqref{equ:cfo_ang013}, is assumed
throughout this section in order to distinctly elaborate on the degrading
effects of a CFO.
\begin{table}[ht]
\caption{Summary of the main PHY parameters of the exemplary UW-OFDM and
  CP-OFDM setup.}
\begin{center}
\begin{scriptsize}
\begin{tabular}{lc|rr} \hline
&& UW-OFDM & CP-OFDM \\ \hline\hline
DFT size & $N$ & 64 & 64 \\ \hline
data subcarriers & $N_d$, $N'_d$ & 32 & 48 \\ \hline
zero subcarriers & $N_z$ & 12 & 12 \\ \hline
pilot subcarriers & $N_p$ & 4 & 4 \\ \hline
red. subcarriers & $N_r$ & 16 & -\\ \hline
guard interval samples & $N_g,N_u$ & 16 & 16\\ \hline
 & & & \\
zero subcarrier indices & $\mathcal{I}_z$ & \{0,27,28,\dots,37\} &\{0,27,28,\dots,37\}\\
 & & & \\
pilot subcarrier indices & $\mathcal{I}_p$ & \{7,21,43,57\} & \{7,21,43,57\}\\
 & & & \\
red. subcarrier indices & $\mathcal{I}_r$ & \{2,5,9,13,17,20,24 &\\
&& 26,38,40,44 & \{\}\\
&& 47,51,54,58,62\}& \\
 & & & \\ \hline
DFT length & $T_\text{DFT}$ &  3.2\,$\mu$s & 3.2\,$\mu$s\\ \hline
guard interval length & $T_\text{GI}$ & 0.8\,$\mu$s & 0.8\,$\mu$s\\ \hline
OFDM symbol lengh & $T_\text{OFDM}$ & 3.2\,$\mu$s & 4\,$\mu$s \\ \hline
subcarrier spacing & $\Delta_f$ & 312.5\,kHz& 312.5\,kHz\\ \hline
\end{tabular}
\end{scriptsize}
\label{tab:setups}
\end{center}
\end{table}
The results are obtained by averaging over $10^4$ independent realizations of channel
impulse responses with unit energy, drawn from
a model with exponentially decaying power delay profile \cite{Fak97} and
assuming a channel delay spread of $\delayspread=100$\,ns.

The CFO $\epsilon$ has been
estimated according to \eqref{equ:cfo_ang032} once per burst based on the estimated CPE of the first OFDM symbol $\hat{\varphi}_0$.
The CPEs $\varphilest$ are estimated
according to \eqref{equ:cfo_ang005} with $\m{W}_p=|\mf{H}_p|^2$. The equalizer is
the same as for CP-OFDM and has been chosen
to be $\m{E}_p=\begin{bmatrix}\m{0} & \m{I} \end{bmatrix}\m{P}_p^T$. Perfect
channel knowledge is assumed at the receiver.

\subsection{Simulation Results}

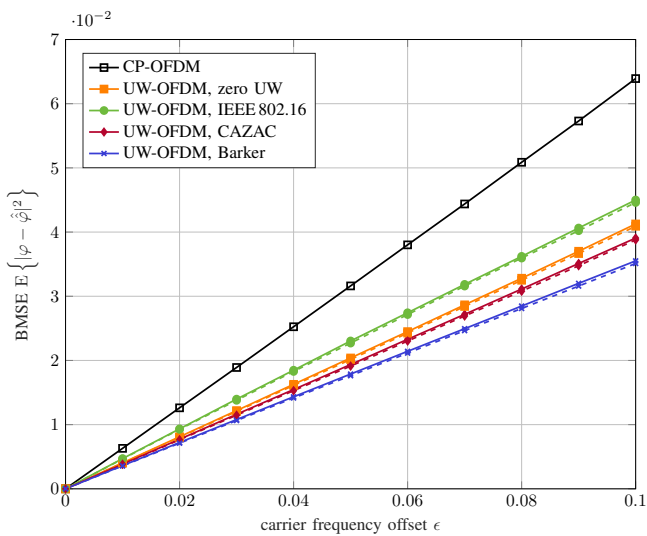
\begin{figure}[!tbh]
  \centering  
  \scalebox{0.66}{
%
\definecolor{mycolor1}{rgb}{1.00000,0.50196,0.00000}%
\definecolor{mycolor2}{rgb}{0.43529,0.72941,0.23922}%
\definecolor{mycolor3}{rgb}{0.70588,0.01961,0.18039}%
\definecolor{mycolor4}{rgb}{0.22353,0.23922,0.83137}%
\begin{tikzpicture}

\begin{axis}[%
width=4.521in,
height=3.566in,
at={(0.758in,0.481in)},
scale only axis,
xmin=0,
xmax=0.1,
xtick={0,0.02,0.04,0.06,0.08,0.1},
xticklabels={{$0$},{$0.02$},{$0.04$},{$0.06$},{$0.08$},{$0.1$}},
xlabel style={font=\color{white!15!black}},
xlabel={carrier frequency offset $\epsilon$},
ymin=0,
ymax=0.07,
ylabel style={font=\color{white!15!black}},
ylabel={BMSE $\E{|\varphi-\hat{\hat{\varphi}}|^2}$},
axis background/.style={fill=white},
xmajorgrids,
ymajorgrids,
legend style={at={(0.03,0.97)}, anchor=north west, legend cell align=left, align=left, draw=white!15!black}
]
\addplot [color=black, line width=1.0pt, mark=square, mark options={solid, black}]
  table[row sep=crcr]{%
0	1.17049352370084e-16\\
0.01	0.00630356939634503\\
0.02	0.0126180916353104\\
0.03	0.0189027691407665\\
0.04	0.0252501526997939\\
0.05	0.0316117679035142\\
0.06	0.0380078758923674\\
0.07	0.0443777066192631\\
0.08	0.0508732057560824\\
0.09	0.057304485760044\\
0.1	0.0639132549692642\\
};
\addlegendentry{CP-OFDM}

\addplot [color=mycolor1, line width=1.0pt, mark=square*, mark options={solid, fill=mycolor1, mycolor1}]
  table[row sep=crcr]{%
0	1.28184717972868e-16\\
0.01	0.0040462894002099\\
0.02	0.00810616913629763\\
0.03	0.0121877398943425\\
0.04	0.0162776947898805\\
0.05	0.0203845297177944\\
0.06	0.0244924665385435\\
0.07	0.0286613595099993\\
0.08	0.0328071972081703\\
0.09	0.037009927705653\\
0.1	0.0412655501743431\\
};
\addlegendentry{UW-OFDM, zero UW}

\addplot [color=mycolor2, line width=1.0pt, mark=*, mark options={solid, fill=mycolor2, mycolor2}]
  table[row sep=crcr]{%
0	1.53677727085467e-16\\
0.01	0.0047035411531655\\
0.02	0.00934155824703333\\
0.03	0.0139716625111569\\
0.04	0.0184789860191239\\
0.05	0.0230356086479055\\
0.06	0.0274283894396075\\
0.07	0.0318503501084168\\
0.08	0.036215263421102\\
0.09	0.0406292897037486\\
0.1	0.0450001176490205\\
};
\addlegendentry{UW-OFDM, IEEE\,802.16}

\addplot [color=mycolor3, line width=1.0pt, mark=diamond*, mark options={solid, fill=mycolor3, mycolor3}]
  table[row sep=crcr]{%
0	1.29162304349863e-16\\
0.01	0.00386095212582865\\
0.02	0.00772407804972243\\
0.03	0.011611401071407\\
0.04	0.0154641501346928\\
0.05	0.0193609369089417\\
0.06	0.023291770088168\\
0.07	0.0271815780607054\\
0.08	0.0311225627973535\\
0.09	0.0351070443549468\\
0.1	0.0390648502663714\\
};
\addlegendentry{UW-OFDM, CAZAC}

\addplot [color=mycolor4, line width=1.0pt, mark=x, mark options={solid, fill=mycolor4, mycolor4}]
  table[row sep=crcr]{%
0	1.23417796249866e-16\\
0.01	0.00361432724402535\\
0.02	0.00721678931654545\\
0.03	0.0107885158625381\\
0.04	0.0143356108962094\\
0.05	0.0178785873512949\\
0.06	0.021419993906551\\
0.07	0.0249438719673728\\
0.08	0.0284732180190347\\
0.09	0.0320107096002409\\
0.1	0.0355121708798968\\
};
\addlegendentry{UW-OFDM, Barker}

\addplot [color=mycolor1, dashed, line width=1.0pt, mark=square*, mark options={solid, fill=mycolor1, mycolor1}, forget plot]
  table[row sep=crcr]{%
0	1.56110291720772e-16\\
0.01	0.00400913585156776\\
0.02	0.00802636430664455\\
0.03	0.012057271335637\\
0.04	0.0161097221192092\\
0.05	0.0201454224960804\\
0.06	0.0242586719997349\\
0.07	0.0284100019988814\\
0.08	0.0324911825870356\\
0.09	0.0366545104790588\\
0.1	0.0409049978943317\\
};
\addplot [color=mycolor2, dashed, line width=1.0pt, mark=*, mark options={solid, fill=mycolor2, mycolor2}, forget plot]
  table[row sep=crcr]{%
0	1.84030310352342e-16\\
0.01	0.0046530779512656\\
0.02	0.00926323664169481\\
0.03	0.0137897089659044\\
0.04	0.0182855397622224\\
0.05	0.0227030229217835\\
0.06	0.0271492103920402\\
0.07	0.031587626447917\\
0.08	0.0359499128949269\\
0.09	0.040213694013334\\
0.1	0.0446072369136795\\
};
\addplot [color=mycolor3, dashed, line width=1.0pt, mark=diamond*, mark options={solid, fill=mycolor3, mycolor3}, forget plot]
  table[row sep=crcr]{%
0	1.58523951500073e-16\\
0.01	0.0038254885000628\\
0.02	0.00764266225371043\\
0.03	0.0114671738805842\\
0.04	0.0152894997974063\\
0.05	0.0191171134048866\\
0.06	0.0230188858489208\\
0.07	0.0269190577036786\\
0.08	0.0307880131101594\\
0.09	0.0347573714720777\\
0.1	0.0387818369106642\\
};
\addplot [color=mycolor4, dashed, line width=1.0pt, mark=x, mark options={solid, fill=mycolor4, mycolor4}, forget plot]
  table[row sep=crcr]{%
0	1.5290749866097e-16\\
0.01	0.0035702705046884\\
0.02	0.00711399051854933\\
0.03	0.010637823589197\\
0.04	0.0141504662349897\\
0.05	0.0176180354485306\\
0.06	0.0211425199988691\\
0.07	0.0246456553244703\\
0.08	0.0280953617510188\\
0.09	0.0315848845075343\\
0.1	0.0351083902244464\\
};
\end{axis}
\end{tikzpicture}%
}
\caption{Comparison of estimation error of the CFO induced CPE
$\varphil$ between UW-OFDM and CP-OFDM
  in terms of Bayesian mean square error (BMSE). Investigated scenario:
  $\m{G}'_d$ (solid line), $\m{G}''_d$ (dashed line), $\m{W}_p=|\mf{H}_p|^2$, and a multipath environment with $\delayspread=100$\,ns.}
\label{fig:phi_est_comparison}
\end{figure}

\begin{figure}[!tbh]
  \centering
  \scalebox{0.66}{
%
\definecolor{mycolor1}{rgb}{1.00000,0.50196,0.00000}%
\begin{tikzpicture}

\begin{axis}[%
width=4.521in,
height=3.566in,
at={(0.758in,0.481in)},
scale only axis,
xmin=0,
xmax=0.1,
xtick={0,0.02,0.04,0.06,0.08,0.1},
xticklabels={{$0$},{$0.02$},{$0.04$},{$0.06$},{$0.08$},{$0.1$}},
xlabel style={font=\color{white!15!black}},
xlabel={carrier frequency offset $\epsilon$},
ymin=0,
ymax=0.035,
ylabel style={font=\color{white!15!black}},
ylabel={average mean power of the ICI},
axis background/.style={fill=white},
xmajorgrids,
ymajorgrids,
legend style={at={(0.03,0.97)}, anchor=north west, legend cell align=left, align=left, draw=white!15!black}
]
\addplot [color=black, line width=1.0pt, mark=square, mark options={solid, black}]
  table[row sep=crcr]{%
0	1.40306013442791e-29\\
0.01	0.000318317457997313\\
0.02	0.00127227620286756\\
0.03	0.00286063928074095\\
0.04	0.0050825435520558\\
0.05	0.00793495796387302\\
0.06	0.0114103306376265\\
0.07	0.0155075028583211\\
0.08	0.0202160826554588\\
0.09	0.0255323231709678\\
0.1	0.0314444476695457\\
};
\addlegendentry{$\sigma_{d_\ici}^2$ CP-OFDM}

\addplot [color=black, line width=1.0pt, mark=triangle, mark options={solid, rotate=270, black}]
  table[row sep=crcr]{%
0	6.08022948330059e-31\\
0.01	2.26655271067022e-06\\
0.02	9.05728799979641e-06\\
0.03	2.03454724221386e-05\\
0.04	3.60866670426199e-05\\
0.05	5.62189021653862e-05\\
0.06	8.06629209933889e-05\\
0.07	0.000109322491262232\\
0.08	0.000142084783624906\\
0.09	0.00017882081530179\\
0.1	0.000219385957254746\\
};
\addlegendentry{$\sigma_{p_\ici}^2$ CP-OFDM}

\addplot [color=mycolor1, line width=1.0pt, mark=square*, mark options={solid, fill=mycolor1, mycolor1}]
  table[row sep=crcr]{%
0	1.4160693877923e-29\\
0.01	0.000186118382660734\\
0.02	0.000744284396522279\\
0.03	0.00167436934201357\\
0.04	0.00297333930019392\\
0.05	0.00464455518192095\\
0.06	0.00668209151263916\\
0.07	0.0090904268192439\\
0.08	0.0118579663445418\\
0.09	0.014991691298631\\
0.1	0.0184805289599166\\
};
\addlegendentry{$\sigma_{d_\ici}^2$ UW-OFDM}

\addplot [color=mycolor1, line width=1.0pt, mark=x, mark options={solid, mycolor1}]
  table[row sep=crcr]{%
0	2.53546283394285e-30\\
0.01	5.79596637416515e-06\\
0.02	2.31634035196767e-05\\
0.03	5.20594554613822e-05\\
0.04	9.242511193051e-05\\
0.05	0.000144185308171793\\
0.06	0.000207249049780875\\
0.07	0.000281509562348337\\
0.08	0.000366844465644963\\
0.09	0.000463115972041415\\
0.1	0.000570171108815056\\
};
\addlegendentry{$\sigma_{p_\ici}^2$ UW-OFDM}

\end{axis}

\begin{axis}[%
width=5.833in,
height=4.375in,
at={(0in,0in)},
scale only axis,
xmin=0,
xmax=1,
ymin=0,
ymax=1,
axis line style={draw=none},
ticks=none,
axis x line*=bottom,
axis y line*=left,
legend style={legend cell align=left, align=left, draw=white!15!black}
]
\end{axis}
\end{tikzpicture}%
}
\caption{Comparison of the average mean power of the ICI experienced by a pilot
  symbol in UW-OFDM and CP-OFDM. The ICI is separated into the data part
and the part induced by the other pilots. Investigated scenario: $\m{G}'_d$ and a multipath environment with $\delayspread=100$\,ns.}
\label{fig:ici_pwr_comparison}
\end{figure}
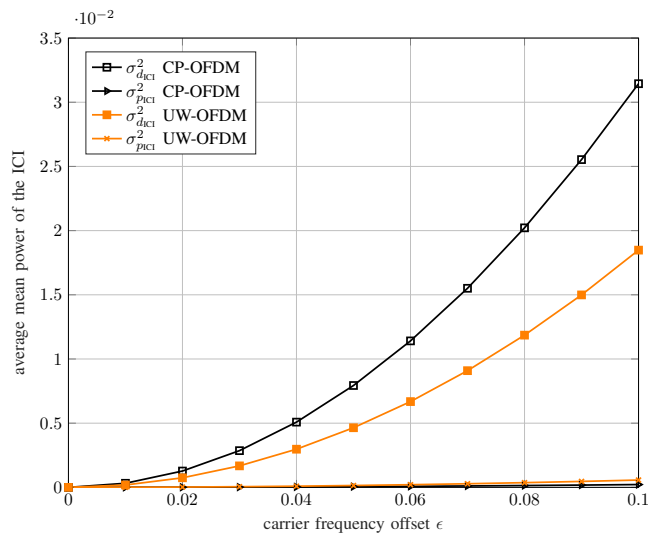

Fig.~\ref{fig:phi_est_comparison} clearly shows that UW-OFDM outperforms
CP-OFDM in terms of estimating $\varphil$ based on frequency pilot tones,
regardless of the utilized generator matrix and the UW.
The performance variation among different UWs is due to two
reasons. First of all, the approximation of $\varphi_{\pil}$ as an affine
function is not equally accurate for all UWs,
cf.~Fig.~\ref{fig:phi_pil_offset}. Second of all, in some cases (e.g., for a
Barker sequence) the portions of the UW
overlaying the pilot subcarriers coincidentally add up coherently with the pilot
symbols and thus increase the resulting
signal-to-interference-noise ratio in the estimation process, cf.~\eqref{equ:cfo_ang022}. The latter is also the reason why a zero UW
--- contrary to intuitive expectations, as in this case there are no UW induced disturbances ---
does not deliver the best results. Independent of that, there is a principle performance gap
between UW-OFDM and CP-OFDM, regardless of a specific UW design. The reason for
this gap becomes immediately apparent when studying all sources degrading
$\hat{\hat{\varphi}}_l$, which can be seen from $\kElempestl$ in
\eqref{equ:cfo_ang016a}. Therefore, let us evaluate the mean power of the data driven ICI disturbances
\begin{align}
\sigma_{d_{\ici}}^2 &= \frac{1}{N_p}\sum_{k=0}^{N_p-1}\E{\kElemdataicil\kElemdataicil^H}\label{equ:cfo_ang034}\\
&=\frac{1}{N_p}\sum_{k=0}^{N_p-1}\E{\ve{e}_k^T\lambdafHl\m{G}_d\datal{\datal}^H\m{G}_d^H\lambdafHl^H(\ve{e}_k^T)^H}\label{equ:cfo_ang035}\\
&=\sigma_d^2\frac{1}{N_p}\sum_{k=0}^{N_p-1}\ve{e}_k^T\mf{\Lambda}_{h,\text{stat}}\m{G}_d\m{G}_d^H\mf{\Lambda}_{h,\text{stat}}^H(\ve{e}_k^T)^H,\label{equ:cfo_ang036}
\end{align}
where $\lambdafHl=\e^{j\varphil}\mf{\Lambda}_{h,\text{stat}}$
and $\kElemdataicil$ denotes the $k$th element of
the vector $\dataicil$ given in \eqref{equ:cfo_ang013a}. It turns out that the
redundancy introduced by the UW-OFDM generator matrix reduces
the resulting interferences significantly, yielding $\sigma_{d_{\ici},\cp}^2 > \sigma_{d_{\ici},\uw}^2 $
for an increasing CFO as shown in Fig.~\ref{fig:ici_pwr_comparison}.
The mean power of the pilot induced ICI
\begin{align} \sigma_{p_{\ici}}^2=\frac{1}{N_p}\sum_{k=0}^{N_p-1}\E{\kElempicil\kElempicil^H}\label{equ:cfo_ang041}
\end{align}
with $\kElempicil=\ve{e}^T_{k}\lambdafHl\sum_{m=0,m\neq
  k}^{N_p-1}\ve{g}_{m}\mElemp$ from \eqref{equ:cfo_ang042} is a little bit higher for
UW-OFDM, but without any relevance. We conclude that UW-OFDM clearly outperforms
CP-OFDM in terms of pilot based estimation of the CFO induced CPE, regardless of the specific setup.


\section{Conclusion}\label{sec:conclusion}
Pilot tones in the frequency domain are a well-established means for estimating
various system parameters. This work has provided a framework to include them
into UW-OFDM signals. The presented framework allows optimizing an UW-OFDM
signal towards two criteria at the same time, to
achieve optimum data estimation capabilities on the one hand, as well as the
most energy efficient pilot
insertion on the other hand. Contrary to conventional OFDM, pilot tone based
CPE estimates incorporate an additional phase offset in case of UW-OFDM, which has to be accounted
for. This phase offset depends on the channel instance and the utilized
setup. An MSE analysis confirms that the same pilot tone based estimation
method provides a significantly better estimate of the CPE in UW-OFDM than in CP-OFDM. This is due the redundancy present in UW-OFDM
signals, decreasing the ICI that causes otherwise a degradation of the
estimation performance.

\appendices
\section{Derivation of CFO model}\label{sec:app_a}
In this Appendix we derive a comprehensive CFO signaling model for UW-OFDM\footnote{Note that parts of the CFO model have
  already been developed in \cite{Kanumalli12}.}. We start with a description of the CFO effects in time domain and finally end up
with the model in \eqref{equ:cfo_ang006}, describing the receive signal of the $l$th
OFDM symbol in the frequency domain.

Assuming a carrier frequency offset $f_\cfo$ present in a system, the
time domain samples in the complex baseband experience an incremental phase shift of
\begin{equation}
y(\idxn T_s)=\e^{j2\pi f_\cfo \idxn T_s}\e^{j\phi_0}x(\idxn T_s)\label{equ:cfo000app},
\end{equation}
where $\idxn$ denotes the discrete time variable, $T_s$ the \underline{s}ampling
time and $\phi_0$ an arbitrary phase offset. Perfect timing
synchronization is expected to take care of $\phi_0$, hence it is set to $\phi_0=0$ and
discarded in the following. The impact of $f_\cfo $ on the OFDM system performance depends on the
relative proportion to
the subcarrier spacing  $\Delta_f$ rather than on the absolute value, thus motivating to introduce a relative carrier frequency offset 
\begin{equation}
\epsilon = \frac{f_\cfo }{\Delta_f}=\frac{f_\cfo }{f_s/N}=\frac{f_\cfo N}{f_s}=f_\cfo NT_s.\label{equ:cfo001app}
\end{equation}
With \eqref{equ:cfo001app} the incremental phase offset in \eqref{equ:cfo000app} translates to
$\e^{j\frac{2\pi \epsilon \idxn}{N}}$ (where $\idxn = 0,1,\dots N-1$ when considering only one OFDM symbol) and
in matrix notation to
\begin{equation}
\ve{y}=\m{\Lambda}'\ve{x}
\end{equation}
with $\m{\Lambda}'\in\mathbb{C}^{N\times N}$ given as $\m{\Lambda}'= \diag{\begin{bmatrix} 1 & \e^{j\frac{2 \pi \epsilon }{N}} & \cdots & \e^{j\frac{2 \pi \epsilon \left(N-1\right) }{N}} \end{bmatrix}^T}$.
In order to take into account the phase accumulated by previous OFDM
symbols and the additional UW in front of the burst (which ensures the cyclic
structure for the first OFDM symbol), a diagonal matrix
$\lambdaNl\in\mathbb{C}^{N\times N}$ with
\begin{equation} 
\lambdaNl  = \e^{j\psil}\m{\Lambda}'=
 \e^{j\frac{2\pi\epsilon\left(Nl+N_{u}\right)}{N}}\m{\Lambda}'
\label{equ:cfo003}
\end{equation}
is introduced, whereas $l\in\{0,1,\dots L-1\}$, $L$
denotes the number of OFDM symbols per burst, and $\psil$ is a phase offset
defined as
\begin{equation}
\psil=\frac{2\pi\epsilon\left(Nl+N_{u}\right)}{N}.\label{equ:psil}
\end{equation}
In the following,
matrices with the notation $'$ as e.g., $\m{\Lambda}'$ encompass the whole
frequency range including the zero subcarriers. The counterpart $\m{\Lambda}$ and similar
matrices introduced later will only comprise non-zero subcarriers.

Starting with
the UW-OFDM transmit signal in \eqref{equ:pil018}
and taking into account \eqref{equ:cfo003}, the $l$th OFDM time domain symbol at the
\underline{r}eceiver can be modelled as
\begin{align}
\Yrxl&= \lambdaNl\m{H}_c\m{F}_N^{-1}\left(\m{B}\m{G}_d\ve{d} +
\m{B}\m{G}_p\ve{p} +\vef{x}_u\right) +\noise\label{equ:cfo004}\\
&= \lambdaNl\m{F}_N^{-1}\m{F}_N\m{H}_c\m{F}_N^{-1}\left(\Xfdownl+ \vef{x}_{p}
+\vef{x}_u\right) +\noise\label{equ:cfo005}\\
&= \lambdaNl\m{F}_N^{-1}\mf{H}'\Xfl + \noise,\label{equ:cfo006}
\end{align}
with $\Xfl = \Xfdownl+ \vef{x}_{p}+\vef{x}_u\in\mathbb{C}^{N\times 1}$
summarizing the effects of \underline{d}ata, \underline{p}ilots and the \underline{U}W in one frequency domain
vector. Applying the DFT yields the
frequency domain OFDM symbol
\begin{align}
\Yfrxl &= \m{F}_N\Yrxl\label{equ:cfo008}\\
&= \m{F}_N\lambdaNl\m{F}_N^{-1}\mf{H}'\Xfl+\m{F}_N\noise\label{equ:cfo009}\\
&= \lambdaNfl\mf{H}'\Xfl + \noiseffull,\label{equ:cfo011}
\end{align}
with a noise vector $\noiseffull\in\mathbb{C}^{N\times 1}$
defined as
\begin{equation}
\noiseffull=\m{F}_N\ve{n}\sim\mathcal{CN}(\ve{0},N\varnoise\m{I}). 
\end{equation}
Since $\lambdaNl$ is diagonal, multiplying with $\m{F}_N$ and $\m{F}_N^{-1}$
results in a circulant matrix $\lambdaNfl$. In order to provide a better
insight on $\lambdaNfl$ and its relationship to $\epsilon$, let us start
with the definition of the $k$th element of vector $\Yfrxl$
\begin{equation}
\kElemYfrxl=\mel{\m{F}_N}_{k,*}\Yrxl =
\sum_{\idxn=0}^{N-1}\e^{-j\frac{2\pi k \idxn}{N}}\nElemYrxl.\label{equ:cfo012}
\end{equation}
According to \eqref{equ:cfo006}, the $\idxn$th element of $\Yrxl$ is expressed as
\begin{align}
\nElemYrxl&= \mel{\lambdaNl}_{\idxn,*}\m{F}_N^{-1}\mf{H}'\Xfl + \nElemNoise\label{equ:cfo013}\\
&= \e^{j\psil}\e^{j\frac{2\pi\epsilon \idxn}{N}}\mel{\m{F}_N^{-1}}_{\idxn,*}\mf{H}'\Xfl+ \nElemNoise\label{equ:cfo014}\\
&= \e^{j\psil}\e^{j\frac{2\pi\epsilon
    \idxn}{N}}\frac{1}{N}\sum_{m=0}^{N-1}\e^{j\frac{2\pi m
    \idxn}{N}}\mel{\mf{H}'}_{m,m}\mElemXfl + \nElemNoise. \label{equ:cfo015}
\end{align}
Note that \eqref{equ:cfo014} follows from \eqref{equ:cfo013} by considering
that only the $u$th element of the vector $\mel{\lambdaNl}_{\idxn,*}$ is non-zero. Plugging \eqref{equ:cfo015} into \eqref{equ:cfo012} and rearranging yields
\begin{align}
\kElemYfrxl &= \sum_{\idxn=0}^{N-1}\e^{-j\frac{2\pi k \idxn}{N}}\e^{j\psil}\e^{j\frac{2\pi\epsilon
    \idxn}{N}}\\
&\times\quad\frac{1}{N}\sum_{m=0}^{N-1}\e^{j\frac{2\pi m \idxn}{N}}\mel{\mf{H}'}_{m,m}\mElemXfl + \kElemVNoisefull\label{equ:cfo016}\\
&= \e^{j\psil}\frac{1}{N}\sum_{\idxn=0}^{N-1}\sum_{m=0}^{N-1}\e^{j\frac{2\pi \left(m
    +\epsilon -k\right)\idxn}{N}}\mel{\mf{H}'}_{m,m}\mElemXfl \nonumber\\
&\quad+ \kElemVNoisefull\label{equ:cfo017}\\
&= \e^{j\psil}\frac{1}{N}\sum_{m=0}^{N-1}\mel{\mf{H}'}_{m,m}\mElemXfl\sum_{\idxn=0}^{N-1}\e^{j\frac{2\pi \left(m
    +\epsilon -k\right)\idxn}{N}}\nonumber\\
&\quad+ \kElemVNoisefull.\label{equ:cfo018}
\end{align}
The relationship between $\Yfrxl$ and $\Xfl$
is fully determined by $\lambdaNfl\mf{H}'$, cf.~\eqref{equ:cfo011}. Applying
this knowledge on \eqref{equ:cfo018} together with
\begin{equation}
\lambdaNfl=
\e^{j\psil}\mf{\Lambda}'\label{equ:cfo100}
\end{equation}
leads to the definition
\begin{align}
\mel{\mf{\Lambda}'}_{k,m}=\frac{1}{N}\sum_{\idxn=0}^{N-1}\e^{j\frac{2\pi}{N}(m+\epsilon-k)\idxn}\qquad
k&=0\dots
N-1;\nonumber\\
m&=0\dots N-1.\label{equ:cfo019}
\end{align}
These formulas allow a compact notation of $\lambdaNfl$, however, an
immediate interpretation of the CFO impact in the frequency domain is rather
difficult. This will thus be provided in the following. Let $\mElemXfhl=\mel{\mf{H}'}_{m,m}\mElemXfl$ for reasons of
compactness, and separate the impact of the subcarrier in consideration
(indicated with index $k$) from
all others, then
\eqref{equ:cfo018} can be rewritten as
\begin{align}
\kElemYfrxl &= \frac{1}{N}\e^{j\psil}\kElemXfhl\sum_{\idxn=0}^{N-1}\e^{j\frac{2\pi\epsilon
    \idxn}{N}}\nonumber\\
&\quad+\frac{1}{N}\e^{j\psil}\sum_{m=0,m\neq
  k}^{N-1}\mElemXfhl\sum_{\idxn=0}^{N-1}\e^{j\frac{2\pi\left(m+\epsilon
    -k\right)\idxn}{N}}  + \kElemVNoisefull\label{equ:cfo021}\\
&=\frac{1}{N}\e^{j\psil}\kElemXfhl\frac{1-\e^{j2\pi\epsilon}}{1-\e^{j\frac{2\pi\epsilon}{N}}}\nonumber\\
&\quad+\frac{1}{N}\e^{j\psil}\sum_{m=0,m\neq
  k}^{N-1}\mElemXfhl\frac{1-\e^{j2\pi\left(m+\epsilon
      -k\right)}}{1-\e^{j\frac{2\pi\left(m+\epsilon -k\right)}{N}}}  + \kElemVNoisefull\label{equ:cfo022}\\
&=
  \frac{1}{N}\e^{j\psil}\frac{\e^{j\pi\epsilon}\left(\e^{-j\pi\epsilon}-\e^{j\pi\epsilon}\right)}{\e^{j\frac{\pi\epsilon}{N}}\left(\e^{-j\frac{\pi\epsilon}{N}}-\e^{j\frac{\pi\epsilon}{N}}\right)}\kElemXfhl
  + \frac{1}{N}\e^{j\psil}\nonumber\\
&\quad\times\sum_{m=0,m\neq
    k}^{N-1}\mElemXfhl\frac{\e^{j\pi\left(m+\epsilon -
      k\right)}}{ \e^{j\frac{\pi\left(m+\epsilon -
      k\right)}{N}}}\nonumber\\
&\qquad\qquad\quad\times\frac{\left(\e^{-j\pi\left(m+\epsilon -
      k\right)}-\e^{j\pi\left(m+\epsilon - k\right)}\right)}{\left(\e^{-j\frac{\pi\left(m+\epsilon -
        k\right)}{N}}-\e^{j\frac{\pi\left(m+\epsilon - k\right)}{N}}\right)}
  + \kElemVNoisefull\label{equ:cfo023}\\
&=\e^{j\psil}\e^{j\frac{\pi\epsilon\left(N-1\right)}{N}}\frac{\sin(\pi\epsilon)}{N\sin(\frac{\pi\epsilon}{N})}\kElemXfhl+ \e^{j\psil}\e^{j\frac{\pi\epsilon(N-1)}{N}}\nonumber\\
  &\quad\times\sum_{m=0,m\neq
    k}^{N-1}\e^{j\frac{\pi(m-k)(N-1)}{N}}\frac{\sin\left(\pi(m+\epsilon-k)\right)}{N\sin\left(\frac{\pi(m+\epsilon-k)}{N}\right)}\mElemXfhl\nonumber\\
  &\quad+ \kElemVNoisefull.\label{equ:cfo024}
\end{align}
Note that \eqref{equ:cfo022} follows from \eqref{equ:cfo021} by applying the formula for the sum of a
geometric series $S_n=\sum_{p=0}^{P-1}r^p=\frac{1-r^{P}}{1-r}$, and \eqref{equ:cfo023} is a preparation step to apply $\left(\e^{-ja} -
\e^{ja}\right)=2j\sin(a)$. Finally, the frequency domain receive signal corrupted by
CFO is
\begin{align}
\kElemYfrxl&= \e^{j\psil}\e^{j\frac{2\pi}{N}\epsilon\left(\frac{N-1}{2}\right)}\frac{\sin(\pi\epsilon)}{N\sin(\frac{\pi\epsilon}{N})}\mel{\mf{H}'}_{k,k}\kElemXfl\nonumber\\
&\quad+ \kElemICI  + \kElemVNoisefull\label{equ:cfo025}\\
&=\e^{j\varphil}\frac{\sin(\pi\epsilon)}{N\sin(\frac{\pi\epsilon}{N})}\mel{\mf{H}'}_{k,k}\kElemXfl
+ \kElemICI  + \kElemVNoisefull,\label{equ:cfo026}
\end{align}
with an ICI term $\kElemICI$ given as
\begin{align}
\kElemICI = \e^{j\varphil}&\sum_{m=0,m\neq
  k}^{N-1}\e^{j\frac{\pi(m-k)(N-1)}{N}}\nonumber\\
&\times\frac{\sin\left(\pi(m+\epsilon-k)\right)}{N\sin\left(\frac{\pi(m+\epsilon-k)}{N}\right)}\mel{\mf{H}'}_{m,m}\mElemXfl,\label{equ:cfo027}
\end{align}
and a phase offset
\begin{equation} \varphil=\psil+\frac{2\pi}{N}\epsilon\left(\frac{N-1}{2}\right)=\frac{2\pi}{N}\epsilon\left(Nl+N_u+\frac{N-1}{2}\right).\label{equ:varphil}
\end{equation}

As can be seen from \eqref{equ:cfo026}, there are three effects on a subcarrier symbol
$\kElemXfl$ as a result of a CFO:
\begin{itemize}
\item A phase offset by $\varphil$,
\item an attenuation by $\frac{\sin(\pi\epsilon)}{N\sin(\pi\epsilon/N)}$, and
\item an ICI term $\kElemICI$ with similar properties as additive noise.
\end{itemize}

In matrix notation, \eqref{equ:cfo026} translates to 
\begin{align}
  \Yfrxl&= \lambdaNfl\mf{H}'\Xfl + \noiseffull \label{equ:cfo028}\\
  &=\e^{j\psil}\mf{\Lambda}'\mf{H}'\Xfl
  + \noiseffull\label{equ:cfo029}\\
  &=\e^{j\psil}\e^{j\frac{2\pi}{N}\epsilon\left(\frac{N-1}{2}\right)}\mf{\Lambda}'_\text{stat}\mf{H}'\Xfl
  + \noiseffull\label{equ:cfo030}\\
 &=\e^{j\varphil}\mf{\Lambda}'_\text{stat}\mf{H}'\Xfl + \noiseffull\label{equ:cfo030a},
\end{align}
taking into account \eqref{equ:cfo100} and
\begin{align}
\mf{\Lambda}' &=\e^{j\frac{2\pi}{N}\epsilon\left(\frac{N-1}{2}\right)}\mf{\Lambda}'_\text{stat},\label{equ:cfo031a}\\
\mel{\mf{\Lambda}'_\text{stat}}_{k,m}&=\frac{\sin\left(\pi(m+\epsilon-k)\right)}{N\sin\left(\frac{\pi(m+\epsilon-k)}{N}\right)}\e^{j\frac{\pi(m-k)(N-1)}{N}}.\label{equ:cfo031}
\end{align}
For the main diagonal entries with $k=m$, \eqref{equ:cfo031} collapses to
$\frac{\sin(\pi\epsilon)}{N\sin(\frac{\pi\epsilon}{N})}$. The nomenclature
\underline{stat}ic within $\mf{\Lambda}'_\text{stat}$ refers to the independence from 
the OFDM symbol index $l$.

So far, the derived model is based on $\Yfrxl$ which incorporates all
subcarriers.  For data estimation performance analysis, it suffices to consider only the
subcarriers carrying the \underline{p}ay\underline{l}oad. We thus discard the zero subcarriers and yield the vector 
\begin{equation}
\Yfdownl = \m{B}^T\Yfrxl.\label{equ:cfo_rx001}
\end{equation}
Excluding zero subcarriers is straightforward in a CFO free case given in \eqref{equ:pil019} due to the diagonal structure of the channel
matrix $\mf{H}$. However, $\lambdaNfl$ in \eqref{equ:cfo011} is a dense matrix, hence
requiring more in-depth investigations. Conducting a few derivations
(for details we refer to \cite{Hofbauer16}), the receive signal in
\eqref{equ:cfo_rx001} can be expressed as
\begin{align}
  \begin{split}
\Yfdownl=& \lambdafl\mf{H}\m{G}_d\datal +
\lambdafl\mf{H}\m{G}_p\ve{p} + \lambdafl\mf{H}\m{B}^T\vef{x}_u\\
&+\mf{\Lambda}_{zn}^{(l)}\mf{H}_z\vef{x}_{u,z} + \noisef.\label{equ:cfo_rx019}
\end{split}
\end{align}
whereas $\mathcal{I}_{nz}=\mathcal{I}_N\backslash\mathcal{I}_z$ denotes the
set of \underline{n}on-\underline{z}ero subcarrier indices,
$\mathcal{I}_N=\{0,\dots,N-1\}$, and $\mathcal{I}_z$ represents the
positions of \underline{z}ero subcarriers. The offset
$\mf{\Lambda}_{zn}^{(l)}\mf{H}_z\vef{x}_{u,z}$ with $\mel{\mf{\Lambda}_{zn}^{(l)}}_{k,m}=\mel{\lambdaNfl}_{\mathcal{I}_{nz}(k),\mathcal{I}_{z}(m)}$ corresponds to that portion of the UW in
frequency domain, which overlays potential \underline{z}ero subcarriers, i.e.,
$\vef{x}_{u,z}$,  and is then spread on
\underline{n}on-zero subcarriers due to ICI. According to
Sec.~\ref{sec:approx_err_rx_model}, we can safely
assume that $\mf{\Lambda}_{zn}^{(l)}\mf{H}_z\vef{x}_{u,z}\rightarrow \ve{0}$, yielding the
final receive model
\begin{equation}
\Yfdownl\approx \lambdafl\mf{H}\m{G}_p\ve{p} + \lambdafl\mf{H}\m{G}_d\datal
 + \lambdafl\mf{H}\m{B}^T\vef{x}_u + \noisef
\end{equation}
with negligible approximation error.

\bibliographystyle{./sty/IEEEtran}
\bibliography{./sty/IEEEabrv,uwofdm}

\end{document}